\documentclass[twocolumn,aps,prb,eqsecnum,floatfix]{revtex4}
\usepackage{graphicx,bm,times,amsmath,multirow}
\allowdisplaybreaks
\newcommand{\I} {\mathrm{I}}
\newcommand{\II}{\mathrm{I\hspace{-.1em}I}}

\begin{document}
 
\title{
\textit{Ab initio} derivation of 
multi-orbital extended Hubbard model 
for molecular crystals
}

\author{Masahisa Tsuchiizu}
\author{Yukiko Omori}
\author{Yoshikazu Suzumura}
\affiliation{
Department of Physics, Nagoya University, Nagoya 464-8602, Japan
}

\author{Marie-Laure Bonnet} 
\affiliation{
Institute of Physical Chemistry,
University of Zurich, Winterthurerstrasse 190,
8057 Zurich, Switzerland 
}

\author{Vincent Robert}
\affiliation{
Universit\'e de Lyon, Laboratoire de Chimie,
Ecole Normale Sup\'erieure de Lyon, 
CNRS, 46 all\'ee d'Italie, F-69364 Lyon, France
}
\affiliation{
Laboratoire de Chimie Quantique,
UMR 7177 CNRS/Universit\'e de Strasbourg,
4, rue Blaise Pascal F-67000 Strasbourg, France
}
\date{January 30, 2012}

\begin{abstract}
From configuration interaction (CI) \textit{ab initio} calculations, 
we derive an effective two-orbital extended Hubbard model
based on the gerade (g) and ungerade (u) molecular orbitals (MOs) 
  of the charge-transfer molecular conductor (TTM-TTP)I$_3$ and
  the single-component molecular conductor [Au(tmdt)$_2$]. 
First, 
by focusing on the isolated molecule, we 
 determine the parameters for the model Hamiltonian
  so as to 
  reproduce the CI Hamiltonian matrix. 
Next, we extend the analysis to 
   two neighboring molecule pairs in the crystal and 
  we perform similar calculations 
   to evaluate the inter-molecular interactions. 
From the resulting tight-binding parameters, 
we analyze the band structure 
  to confirm that two bands overlap and mix in together,  supporting
 the multi-band feature.
Furthermore,
using a fragment decomposition,
we derive the effective model based on the fragment 
MOs and  show that 
  the staking TTM-TTP molecules can be described by
  the zig-zag two-leg ladder with the inter-molecular transfer integral
    being larger than  the intra-fragment transfer integral
  within the molecule.
The inter-site interactions between 
the fragments  
 follow a Coulomb  law,  
supporting the fragment decomposition strategy.
\end{abstract}


\maketitle

\section{Introduction}

In the long history of research on molecular solids, 
 various materials have been synthesized and 
a rich variety of phenomena have been discovered, 
  e.g., Peierls insulator, Mott insulator, charge-ordered state,
   antiferromagnetic state, spin and charge 
  density wave state, and superconducting state.\cite{Seo2004,Seo2006}
For the description of almost all these phases, 
 it has been recognized 
 that
  only one  frontier orbital, the highest-occupied-molecular orbital
  (HOMO) or lowest-unoccupied-molecular orbital (LUMO) plays 
  a crucial role.
\cite{Mori2004,Mori1984_Huckel}
The tight-binding approach, where a molecule 
  is regarded as a single site and the various ways of 
  molecular packing are reflected by the anisotropy 
  of inter-site transfer integrals and of inter-site Coulomb repulsions,
  has been successful in the systematic understanding 
  of the origin of various phases.

Recently theoretical efforts have been devoted to accurate
derivation  of 
 the single-orbital
  Hubbard Hamiltonian from density-functional theory 
(DFT; Refs.\ \onlinecite{Cano-Cortes2007,Scriven2009a,Scriven2009b,Nakamura2009,Kandpal:2009cr,Imada2010})
 and
 wavefunction-based
calculations.
\cite{Calzado2002a,Calzado2002b,Verot2011}
  The magnitude of the bare ``on-site'' Coulomb interaction  
  within the same molecular orbital (MO) of the benchmark TTF molecule
  was  evaluated   as $\sim$
  5.9 eV,  which is reduced to 4.7 eV by taking into account 
  the intra-molecular screening 
effects. \cite{Cano-Cortes2007}
The on-site Coulomb repulsion is further reduced
for larger molecules, 
 as expected, e.g.,  in  BEDT-TTF molecule \cite{Scriven2009a,Scriven2009b}
where  it is 
estimated $\sim 4.2$
  eV.
It is generally expected that the intra-molecular
  screening
effects are not so pronounced 
 for  larger molecules 
  due to the nature of delocalized molecular
  orbitals. \cite{Cano-Cortes2007,Verot2011}
The inter-molecular screening effects 
have also been analyzed by 
several approaches.
\cite{Cano-Cortes2007,Scriven2009a,Scriven2009b,Nakamura2009,Imada2010}

Multi-molecular-orbital properties have attracted 
  recent attention in the 
single-component molecular solids,
\cite{Kobayashi2001,Kobayashi2004,Kobayashi2006}
 such as [Ni(tmdt)$_2$] 
 and [Au(tmdt)$_2$]. 
Among these single-component molecular solids,
the metallic behavior was first confirmed 
 in  [Ni(tmdt)$_2$].\cite{Tanaka_SCMM2001,Tanaka2004}
In [Au(tmdt)$_2$],  antiferromagnetic ordering was observed with 
  high transition temperature $T_\mathrm{AF}=110$ K.
   \cite{Suzuki2003,Zhou2006,Hara2008}
Interestingly, it has been proposed  that in [Au(tmdt)$_2$], 
  a new type of antiferromagnetic state, called 
the \textit{intra-molecular} antiferromagnetic (IAF) state,
is realized,  in which
two ligands within each molecule have opposite spins.
\cite{Ishibashi2005,Ishibashi2008}
From DFT-based \textit{ab initio} calculations  
for these compounds,
several MOs are shown to 
 contribute to the conduction band of the crystal systems.
\cite{Ishibashi2005,Ishibashi2008,Rovira}
It has been recognized that 
  the frontier orbitals for these compounds can be decomposed into 
several moieties, i.e., 
 fragment  MOs, and
  the electronic states have been analyzed using
   the microscopic tight-binding Hamiltonian 
based on the fragment MOs.
  \cite{Seo2008}

 The charge-transfer molecular compound  (TTM-TTP)I$_3$ 
(Refs.\ \onlinecite{Mori2004} and \onlinecite{Mori1994,Mori1997,Yasuzuka2006})
is now considered
as another candidate of multi-molecular-orbital system.
  \cite{Marie2010,Tsuchiizu2011_Lett}
In this compound, 
  the non-magnetic insulating 
  behavior at low temperature has been confirmed.
 \cite{Maesato1999,Fujimura1999,Onuki2001_Synth,Onuki2001_JPCS}
A charge ordering reflected by the alternation of valence of
  TTM-TTP molecule 
   along the stacking direction has been proposed 
  initially.\cite{Onuki2001_Synth,Onuki2001_JPCS}
The experimental analysis based on 
  Raman-scattering \cite{Yakushi2003,Swietlik2004,Swietlik2005} and  
  x-ray \cite{Nogami2003} measurements suggested
 a new type of charge-ordered state, 
 ``intra-molecular charge ordering (ICO)'', 
  which cannot be described by the conventional single-orbital 
   approximation.
By  performing wavefunction-based
   \textit{ab initio} calculations for the isolated 
 ionic TTM-TTP molecule, \cite{Marie2010} 
 we previously revealed that this system has a 
multiconfigurational character.

In the present paper, we propose a scheme 
to build up an
effective two-orbital extended Hubbard model for 
(TTM-TTP)I$_3$ and [Au(tmdt)$_2$],
from \textit{ab initio} multi-reference
 configuration-interaction (MR-CI) calculations.
 \cite{QuantumChemistry} 
To the best of the authors' knowledge, this is
 the first analysis of parameter evaluations for multi-MO-based
 systems.  
The band structure in the so-called normal phase, 
  i.e., no long-range-ordered state, 
is examined in terms of the evaluated tight-binding parameters.
 By noting that the resulting
 MOs 
exhibit bonding and anti-bonding character
 between the left and right moieties, 
we transform
 the two-orbital model into the fragment MO picture.

Some of the results have already been presented in Ref.\ 
\onlinecite{Tsuchiizu2011_Lett},
 where 
the  band structure for (TTM-TTP)I$_3$
turned out to be consistent with 
  the direct band calculation based
  on the DFT results.
Our goal in the present paper is to give
 an explicit derivation to extract model parameters
 in  (TTM-TTP)I$_3$ and  [Au(tmdt)$_2$] compounds.
 Indeed, one would like to unravel the origins
 of the ICO state in the former and the IAF state in the latter.
The present paper is organized as follows.
In section II, the characteristic feature of MOs
 is briefly reviewed.
In section III, we derive the effective two-orbital 
Hubbard Hamiltonian 
 for the isolated (TTM-TTP)$^+$ 
and [Au(tmdt)$_2$] molecules, and evaluate the parameters
 by using 
\textit{ab initio} calculations.
In section IV, the inter-molecular interactions are evaluated by 
  focusing on  two neighboring molecules in the crystal.
In section V, we examine
the band structure of the crystals.
Section VI is devoted to the application to the 
fragment decomposition of MOs and 
examinations of its validity.

\section{Molecular orbitals}

\begin{figure}[t]
\includegraphics[width=8cm,bb= 0 0 366 205]{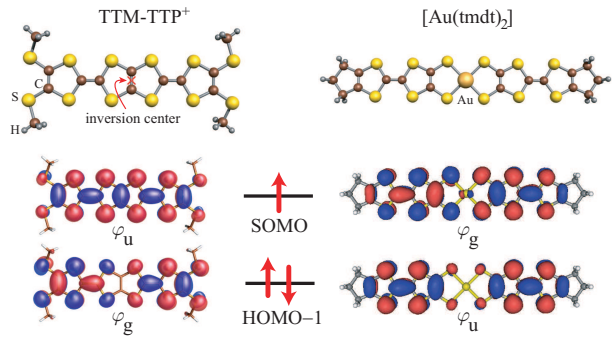}
\caption{
The molecular structures and quasi-degenerate molecular orbitals for
 (TTM-TTP)$^+$ (left) and [Au(tmdt)$_2$] (right).
For (TTM-TTP)$^+$, 
the singly-occupied-molecular-orbital (SOMO) has 
ungerade (u) symmetry with respect to the inversion center, while 
the second-highest-occupied-molecular-orbital (HOMO$-1$),
 has gerade (g) symmetry.
For [Au(tmdt)$_2$],
 the inversion center is located on the metal (Au) center, 
and the SOMOs (HOMO$-1$) are of g (u) symmetries.
}
\label{fig:MO}
\end{figure}

TTM-TTP stands for
2,5-bis[4,5-bis(methylthio)-1,3-dithiol-2-ylidene]-1,3,4,6-tetrathiapentalene
and 
tmdt stands for 
trimethylenetetrathiafulvalenedithiolate.
The molecular structures are
 shown in Fig.\ \ref{fig:MO}.
The crystal structure of  (TTM-TTP)I$_3$ is triclinic and the space group 
  is P$\bar{1}$: the inversion center is located 
  on the mid point of the TTM-TTP molecule.
The formal charge of the TTM-TTP molecule is    $+1$ 
due to the presence of mono-counterion I$_3^-$, and
the HOMO is singly occupied, that is SOMO (singly-occupied-MO) in nature.
For [Au(tmdt)$_2$], the crystal structure is also triclinic and the
  space group is P$\bar{1}$, where
the inversion center is located on the metal center.
Note that the isolated neutral molecule
[Au(tmdt)$_2$]  is a radical species.
All our \textit{ab initio} calculations were
performed using the MOLCAS 7 package. \cite{MOLCAS} 
The atomic coordinates are read from the data of x-ray structure
analysis.
 \cite{Mori1994,Suzuki2003}
Due to the presence of inversion center at the mid point of the 
  molecules, the resulting MOs can be classified into 
   gerade (g) or ungerade (u) with 
   respect to the inversion center.
The energy spectrum was calculated 
  using a restricted open-shell Hartree-Fock (ROHF) procedure. 
Along these ROHF calculations, three electrons 
are likely to occupy the frontier g and u MOs.
These MOs are likely to be singly 
occupied or doubly occupied in the ROHF calculations. 
In order to  estimate  the energies of the expected
levels, we set the occupation numbers for both MOs as 1.5.\cite{Marie2010}
Under these conditions, 
for (TTM-TTP)$^+$, we obtain that the SOMO is u and HOMO$-1$ is g, 
  in agreement with extended H\"uckel calculations, 
\cite{Mori1984_Huckel} and 
the resulting energy difference obtained from the ROHF calculation is 
  $\sim 0.4$ eV.\cite{Marie2010}
For [Au(tmdt)$_2$], we obtain that the SOMO is g and HOMO$-1$ is u,
  in agreement with  DFT-based \textit{ab initio} calculations.
\cite{Ishibashi2005}
The notations for label of MOs are adopted to those in 
Ref.\ \onlinecite{Ishibashi2005}.
The energy difference obtained from the ROHF calculation for isolated
molecule is given by
$(\varepsilon_\mathrm{SOMO}-\varepsilon_\mathrm{HOMO-1})\simeq 0.26$ eV.
Incidentally, the LUMO and the
third-highest-occupied-molecular-orbital (HOMO$-2$) 
 are sufficiently separated in energy, 
$(\varepsilon_\mathrm{LUMO}-\varepsilon_\mathrm{SOMO})\simeq 6.07$ eV and
$(\varepsilon_\mathrm{SOMO}-\varepsilon_\mathrm{HOMO-2})\simeq 2.03$ eV.

\section{Intra-molecular interactions}

We shall now focus on g and u MOs, shown in Fig.\ \ref{fig:MO}, 
 and derive the effective two-orbital Hubbard Hamiltonian
  for the isolated TTM-TTP$^+$ 
and [Au(tmdt)$_2$] molecules.
All the model parameters are determined so as to  reproduce the 
  energies of different electronic configurations
   obtained from the \textit{ab initio} calculations.

\subsection{Construction of the two-orbital Hubbard Hamiltonian}

In this subsection, we derive the expression of the
  two-orbital tight-binding Hamiltonian for the isolated molecule.
The relevant orbitals are written as
  $\varphi_\mathrm{g}(\bm r)$ and
  $\varphi_\mathrm{u}(\bm r)$, respectively.
Due to the inversion symmetry around the 
midpoint of the molecule, the wavefunctions obey
$\varphi_{\mathrm{g}}(\bm{r}) = \varphi_\mathrm{\mathrm{g}}(-\bm{r})$ and
$\varphi_{\mathrm{u}}(\bm{r}) = -\varphi_{\mathrm{u}}(-\bm{r})$,
where the origin of the coordinate is taken as the inversion center
 (Fig.\ \ref{fig:MO}).

By focusing on the frontier orbitals, SOMO and HOMO$-1$, the
full Hamiltonian 
for the  isolated  molecules is expressed as
\begin{eqnarray}
H_{1\mbox{-}\mathrm{mol}}
&=&
\sum_\alpha\sum_\sigma \bar\varepsilon^0_\alpha 
c_{\alpha,\sigma}^\dagger c_{\alpha,\sigma}^{}
\nonumber \\ && {}
+
\smash[b]{ \frac{1}{2} 
\sum_{\alpha,\beta,\alpha',\beta'} 
}
\sum_{\sigma,\sigma'}
(\alpha \alpha' , \beta \beta') \, 
c_{\alpha,\sigma}^\dagger 
c_{\alpha',\sigma}	
c_{\beta,\sigma'}^\dagger 
c_{\beta',\sigma'} 
\nonumber \\ && {}
+ \mathrm{const.},
\label{eq:H_1mol-bare}
\end{eqnarray}
where  $c_{\alpha,\sigma}^\dagger$ stands for the creation operator
of electron
  in MO $\alpha(=\mathrm{g, u})$ with spin $\sigma(=\uparrow,\downarrow)$.
The orbital energies for the isolated molecules 
  are represented by $\bar\varepsilon^0_\alpha$.
For the electron-electron interactions, 
we adopt the
  notation 
  for the integrals as 
\begin{eqnarray}
(\alpha \alpha' , \beta \beta') 
&=&
\iint d{\bm r}_1 d{\bm r}_2
\nonumber \\ &\times& 
\varphi_\alpha^* ({\bm r}_1) 
\varphi_{\alpha'}({\bm r}_1)
\frac{1}{|{\bm r}_1-{\bm r}_2|}
\varphi_\beta^* ({\bm r}_2)
\varphi_{\beta'}   ({\bm r}_2) ,
\label{eq:integral}
\end{eqnarray}
where all the wavefunctions are real.
Under the permutation of ${\bm r}_1$ and ${\bm r}_2$, 
 we have the relations:
\begin{eqnarray}
(\alpha \alpha' , \beta \beta')
 &=& 
( \beta \beta', \alpha \alpha')
 = 
(\alpha \alpha' , \beta' \beta)
\nonumber \\ 
 &=& 
(\alpha' \alpha , \beta \beta')
 = 
(\alpha' \alpha , \beta' \beta).
\label{eq:relation}
\end{eqnarray}
Due to the g/u symmetry of the MOs and 
  relations  (\ref{eq:relation}),
  the number of 
  independent interactions can be  reduced.
 Depending on the choice of  
the indices $\alpha, \beta, \alpha', \beta'$ in Eq.\ (\ref{eq:integral}),
  we can classify those integrals
  into  intra- and inter-orbital interactions.
The intra-orbital interactions 
$( \alpha\alpha , \alpha\alpha )$ 
represents
  the magnitude of interaction for electrons within the same MOs, 
  in other words, the ``on-site'' Coulomb repulsions.
Here 
 we define the parameters:
\begin{eqnarray}
U_{\mathrm{g}} 
=
(\mathrm g \mathrm g , \mathrm g \mathrm g )
,\quad
U_{\mathrm{u}} 
=
( \mathrm u \mathrm u , \mathrm u \mathrm u ).
\label{eq:UgUu}
\end{eqnarray}
For the inter-orbital case, the interactions involving odd number of
  ungerade MO, e.g., 
   $( \mathrm g \mathrm g ,\mathrm g \mathrm u)$
  vanishes due to the symmetry constraint. 
The possible interactions are
  $(\alpha\alpha, \beta\beta )$,
  $(\alpha\beta , \beta\alpha )$, and
  $( \alpha\beta , \alpha\beta )$
with $\alpha\neq\beta$, 
 which correspond to 
 the Coulomb integral, 
  exchange integral, and pair-hopping interaction, 
  respectively.
We note that from Eq.\ (\ref{eq:relation}), 
  the amplitudes of 
  the exchange integral and the pair-hopping interaction become identical
 $( \alpha \beta , \beta \alpha ) =
  ( \alpha \beta , \alpha \beta )$.
We define the two-independent coupling constants given by 
\begin{subequations}
\begin{eqnarray}
J_\mathrm{H} &=&   
   2 ( \mathrm{gu} , \mathrm{ug} )
=  2 ( \mathrm{gu} , \mathrm{gu} ),
\\ \nonumber \\
U' &=& 
 ( \mathrm{gg} , \mathrm{uu} )
 - \frac{1}{2} ( \mathrm{gu} , \mathrm{ug} ),
\end{eqnarray}%
\label{eq:JU'}%
\end{subequations}
where $J_\mathrm{H}$ represents the Hund exchange coupling 
  including the pair-hopping term, and 
 $U'$ is the inter-orbital Coulomb repulsion.

The level energies 
  are determined by the eigenvalues of the ROHF.
We note that the level energies $\bar\epsilon^0_\alpha$ 
 in Eq.\ (\ref{eq:H_1mol-bare})
do not
correspond to the eigenvalues of ROHF,
  since the Hartree contributions arising from electron-electron
  interactions are not taken into account.
These effects can be incorporated in terms of the model parameters 
  $U_{\mathrm{g}}$, $U_{\mathrm{u}}$, and $U'$,
 and 
  we can re-define the MO level energies by  
\begin{subequations}
\begin{eqnarray}
\varepsilon_\mathrm{g}^0 
&=& 
\bar\varepsilon_\mathrm{g}^0 
+ U_\mathrm{g}\langle c_{\mathrm{g},\uparrow}^\dagger 
                      c_{\mathrm{g},\uparrow}^{}  \rangle
 + U' \sum_\sigma \langle c_{\mathrm{u},\sigma}^\dagger 
                      c_{\mathrm{u},\sigma}^{} \rangle,
\\ \nonumber \\
\varepsilon_\mathrm{u}^0 
&=&
\bar\varepsilon_\mathrm{u}^0 
+ U_\mathrm{u}\langle c_{\mathrm{u},\uparrow}^\dagger 
                      c_{\mathrm{u},\uparrow}^{}  \rangle
 + U' \sum_\sigma \langle c_{\mathrm{g},\sigma}^\dagger 
                      c_{\mathrm{g},\sigma}^{} \rangle,
\end{eqnarray}%
\label{eq:epsilon0}%
\end{subequations}
where the expectation values are
 determined by the Hartree-Fock calculation.
Since the average charges
  on the each MO were set to $3/2$,
 \cite{Marie2010} 
 we obtain 
$\varepsilon_\mathrm{g}^0 
=
\bar\varepsilon_\mathrm{g}^0 + \frac{3}{4}U_\mathrm{g} + \frac{3}{2}U'$ and
$\varepsilon_\mathrm{u}^0 
= 
\bar\varepsilon_\mathrm{u}^0 + \frac{3}{4}U_\mathrm{u} + \frac{3}{2}U'$.

In order to express the Hamiltonian in 
  a compact form, 
 we introduce the density operators 
 in the normal-ordered form:
\begin{eqnarray}
n_{\mathrm{g},\sigma}
=
c_{\mathrm{g},\sigma}^\dagger c_{\mathrm{g},\sigma}^{}-\frac{3}{4}
, 
\quad
n_{\mathrm{u},\sigma}
=
c_{\mathrm{u},\sigma}^\dagger c_{\mathrm{u},\sigma}^{}-\frac{3}{4},
\label{eq:densityop}
\end{eqnarray}
and
$n_{\mathrm{g}}=(n_{\mathrm{g},\uparrow}+n_{\mathrm{g},\downarrow})$,
$n_{\mathrm{u}}=(n_{\mathrm{u},\uparrow}+n_{\mathrm{u},\downarrow})$.
The Coulomb interactions having amplitudes
  $U_{\mathrm{g}}$, $U_{\mathrm{u}}$, and $U'$ can be expressed 
  in bilinear forms of  these density operators.
In terms of model parameters defined 
 in Eqs.\ (\ref{eq:UgUu}), (\ref{eq:JU'}), and 
(\ref{eq:epsilon0}), and also in terms of 
  density operators given in Eq.\ (\ref{eq:densityop}), 
we obtain
the effective two-orbital tight-binding Hamiltonian for the isolated
  molecules:
\begin{eqnarray}
H_{1\mbox{-}\mathrm{mol}} 
&=& 
E_0 
+ \varepsilon_{\mathrm g}^0 n_{\mathrm g}
+ \varepsilon_{\mathrm u}^0 n_{\mathrm u} 
\nonumber \\ && {} 
 + U_{\mathrm g} n_{\mathrm g\uparrow} n_{\mathrm g\downarrow}
 + U_{\mathrm u} n_{\mathrm u\uparrow} n_{\mathrm u\downarrow}
 + U' n_{\mathrm g} n_{\mathrm u}
\nonumber \\ && {} 
-J_{\mathrm H} 
\left[
\bm{S}_{\mathrm g} \cdot \bm{S}_{\mathrm u}
-\frac{1}{2}
 (c_{\mathrm{g},\uparrow}^\dagger c_{\mathrm{u},\uparrow}^{} 
  c_{\mathrm{g},\downarrow}^\dagger c_{\mathrm{u},\downarrow}^{}
   + \mathrm{h.c.})
\right],
\nonumber \\
\label{eq:singlemolecule}
\end{eqnarray}
where $\bm{S}_\alpha$ is the spin operator given by
  $\bm{S}_\alpha=\frac{1}{2}\sum_{\sigma,\sigma'}c_{\alpha,\sigma}^\dagger 
 \bm{\sigma}_{\sigma,\sigma'} c_{\alpha,\sigma'}$ 
  with $\bm{\sigma}$ being the Pauli matrix.
The energy constant $E_0$ and the level energy 
 $\varepsilon_\mathrm{g}^0$ ($\varepsilon_\mathrm{u}^0$)
correspond to
 the total energy and the MO level energy for $\varphi_\mathrm{g}$ ($\varphi_\mathrm{u}$),
obtained by the ROHF calculations, respectively.

\subsection{Possible configurations}

In this subsection, we introduce the possible configurations
    for the two-orbital system
 including 0, 1, 2, 3, or 4 electrons.
Let us stress that the MOs used to describe the different states remain
unchanged and correspond to the three-electron case.
For this purpose, we introduce 
the following notation:
\begin{eqnarray}
\left| 
      \begin{array}{c} 
\displaystyle
         \frac{\mbox{\small \, u \,}}{} 
\\
\displaystyle
         \frac{\mbox{\small \, g \,}}{} 
      \end{array}
\right\rangle
.
\end{eqnarray}

\begin{subequations}\label{eq:configurations-1mol}

The zero-electron state is given by
\begin{eqnarray}
| \psi^{(0)} \rangle &=&
\left| 
  \begin{array}{c} 
   - \\
   -
  \end{array}
\right\rangle
= 
| 0 \rangle,
\end{eqnarray}
where $|0\rangle $ represents the Slater determinant 
  in which the MOs are occupied up to 
the third-highest-occupied-MO.

For the one-electron states, we have two configurations 
\begin{eqnarray}
| \psi^{(1)}_{\mathrm{g}} \rangle &=&
\left| 
\begin{array}{c} 
 - \\
\, \uparrow \,
\end{array}
\right\rangle
=
c_{\mathrm{g},\uparrow}^\dagger 
| 0 \rangle
,
\\ \nonumber \\
| \psi^{(1)}_{\mathrm{u}} \rangle &=&
\left| 
\begin{array}{c} 
\, \uparrow \, \\
-
\end{array}
\right\rangle
=
c_{\mathrm{u},\uparrow}^\dagger 
| 0 \rangle
.
\end{eqnarray}
Since the 
  same relations hold for the spin-down states,
we focus only on the spin-up states hereafter.

For the two-electron states, there are four possible
configurations, which can be classified depending on the 
  symmetry and spin states.
For the two-electron 
states with gerade symmetry, 
  we have two configurations:
\begin{eqnarray}
|\psi^{(2)}_{\mathrm{g,s,1}}\rangle &=&
\left| 
\begin{array}{c} 
 - \\
\uparrow\downarrow 
\end{array}
\right\rangle
=
c_{\mathrm{g},\uparrow}^\dagger c_{\mathrm{g},\downarrow}^\dagger 
| 0 \rangle
,
\label{eq:2-el-gs1}
\\ \nonumber \\
| \psi^{(2)}_{\mathrm{g,s,2}} \rangle &=&
\left| 
\begin{array}{c} 
\uparrow\downarrow \\
-
\end{array}
\right\rangle
=
c_{\mathrm{u},\uparrow}^\dagger c_{\mathrm{u},\downarrow}^\dagger 
| 0 \rangle
,
\label{eq:2-el-gs2}
\end{eqnarray}
where both configurations are spin singlets.
For the two-electron 
states
 with ungerade symmetry, 
there are also  two configurations:
\begin{eqnarray}
|\psi^{(2)}_{\mathrm{u,s}} \rangle &=&
\left| 
  \begin{array}{c} 
  \, \downarrow \, \\
  \, \uparrow  \,
  \end{array}
\right\rangle
=
\frac{1}{\sqrt{2}}
(
c_{\mathrm{g},\uparrow}^\dagger c_{\mathrm{u},\downarrow}^\dagger 
+
c_{\mathrm{u},\uparrow}^\dagger c_{\mathrm{g},\downarrow}^\dagger 
)
| 0 \rangle
,
\qquad
\\ \nonumber \\
| \psi^{(2)}_{\mathrm{u,t}} \rangle &=&
\left| 
  \begin{array}{c} 
  \, \uparrow \, \\ \, \uparrow \,
  \end{array}
\right\rangle
=
c_{\mathrm{g},\uparrow}^\dagger c_{\mathrm{u},\uparrow}^\dagger 
| 0 \rangle
,
\end{eqnarray}
where the 
suffixes
 s and t represent the spin-singlet and spin-triplet
states, respectively. The $S^z=1$ component is only considered
for the spin-triplet state.

The three-electron states are given by 
\begin{eqnarray}
| \psi^{(3)}_{\mathrm{g}} \rangle &=& 
\left| 
\begin{array}{cc}
 \uparrow\downarrow \\
\, \uparrow \, 
\end{array}
\right\rangle
=
c_{\mathrm{g},\uparrow}^\dagger c_{\mathrm{u},\uparrow}^\dagger 
c_{\mathrm{u},\downarrow}^\dagger 
| 0 \rangle
,
\\ \nonumber \\
| \psi^{(3)}_{\mathrm{u}} \rangle &=& 
\left| 
  \begin{array}{c} 
  \, \uparrow \,  \\
  \uparrow\downarrow 
  \end{array}
\right\rangle
=
c_{\mathrm{g},\uparrow}^\dagger c_{\mathrm{u},\uparrow}^\dagger 
c_{\mathrm{g},\downarrow}^\dagger 
| 0 \rangle
.
\end{eqnarray}

Finally the four-electron state has a unique configuration:
\begin{eqnarray}
| \psi^{(4)} \rangle &=&
\left| 
  \begin{array}{c} 
  \uparrow\downarrow \\
  \uparrow\downarrow
  \end{array}
\right\rangle
=
c_{\mathrm{g},\uparrow}^\dagger c_{\mathrm{u},\uparrow}^\dagger 
c_{\mathrm{g},\downarrow}^\dagger c_{\mathrm{u},\downarrow}^\dagger 
| 0 \rangle.
\end{eqnarray}%
\end{subequations}%

\subsection{\textit{Ab initio} calculations}

The energies 
of the different configurations were evaluated 
 by performing
MR-CI
 calculations.\cite{QuantumChemistry} 
Several basis sets were used for convergence control, however,
a weak dependence on the choice of basis
 set is observed.
Throughout this paper, we adopt 
the basis sets
contractions for the elements S(7s6p1d)/[4s3p1d], C(5s5p1d)/[3s2p1d],
Au(13s10p9d6f)/[5s4p4d2f],
and H(3s)/[1s].
The SOMO and HOMO$-1$ levels 
are well separated from the other MOs. 
Therefore, 
these two MOs are used to generate 
a so-called model space
 containing three electrons in two MOs.
In order to evaluate full parameters, we also consider virtual states 
  by removing/adding electrons.

We note that the two-electron
 gerade wavefunctions are given by the superpositions of the
two configurations [Eqs.\ (\ref{eq:2-el-gs1}) and (\ref{eq:2-el-gs2})].
By taking advantage of the information conveyed by the wavefunctions, 
 we can access the Hamiltonian matrix expressed in the basis of
 Eqs.\ (\ref{eq:2-el-gs1}) and (\ref{eq:2-el-gs2}).
Detailed formulation is given later (see Sec.\ \ref{abinitio-2mol}).

\subsection{Parameter mapping}

The model parameters 
  in Eq.\ (\ref{eq:singlemolecule}) can now be evaluated 
 by relating 
  the \textit{ab initio} calculation results 
  with the respective energies expressed in terms of the 
  microscopic model parameters.
From the Hamiltonian Eq.\ (\ref{eq:singlemolecule}) and 
  the configurations [Eq.\ (\ref{eq:configurations-1mol})], 
the energy expectation for the zero-electron state
$E^{(0)}=\langle \psi^{(0)} |H_\mathrm{1\mbox{-}mol}| \psi^{(0)}\rangle$
can be expressed as
\begin{eqnarray}
E^{(0)} =
E_0
 - \frac{3}{2} \varepsilon^0_{\mathrm g} 
 - \frac{3}{2} \varepsilon^0_{\mathrm u}
 + \frac{9}{16} U_{\mathrm g}
 + \frac{9}{16} U_{\mathrm u} 
 + \frac{9}{4} U' 
.
\end{eqnarray}
In the similar way, the respective energies 
$E^{(i)}
  =\langle \psi^{(i)} |H_\mathrm{1\mbox{-}mol}|
  \psi^{(i)}\rangle$
  can be expressed as
\begin{subequations}
\begin{eqnarray}
E_{\mathrm{g}}^{(1)} 
&=&
E^{(0)} 
 + \varepsilon^0_{\mathrm g} 
 - \frac{3}{4} U_{\mathrm g}
 - \frac{3}{2} U' 
,
\\ \nonumber \\
E_{\mathrm{u}}^{(1)} 
&=&
E^{(0)} 
 + \varepsilon^0_{\mathrm u}
 - \frac{3}{4} U_{\mathrm u} 
 - \frac{3}{2} U' 
,
\\ \nonumber \\
E_{\mathrm{g,s,1}}^{(2)} 
&=& 
E^{(0)} 
 + 2 \varepsilon^0_{\mathrm g} 
 - \frac{1}{2} U_{\mathrm g}
 - 3 U' 
,
\\ \nonumber \\
E_{\mathrm{g,s,2}}^{(2)} 
&=&
E^{(0)} 
 + 2 \varepsilon^0_{\mathrm u}
 - \frac{1}{2} U_{\mathrm u} 
 - 3 U' 
,
\\ \nonumber \\
E^{(2)}_{\mathrm{u,s}} &=& 
E^{(0)} 
 + \varepsilon^0_{\mathrm g} 
 + \varepsilon^0_{\mathrm u}
 - \frac{3}{4} U_{\mathrm g}
 - \frac{3}{4} U_{\mathrm u} 
 - 2 U' 
 + \frac{3}{4} J_{\mathrm H} 
,
\nonumber \\
\\ 
E^{(2)}_{\mathrm{u,t}} &=& 
E^{(0)} 
 + \varepsilon^0_{\mathrm g} 
 + \varepsilon^0_{\mathrm u}
 - \frac{3}{4} U_{\mathrm g}
 - \frac{3}{4} U_{\mathrm u} 
 - 2 U' 
 - \frac{1}{4} J_{\mathrm H} 
,
\nonumber \\
\\ 
 E^{(3)}_{\mathrm{g}} &=& 
E^{(0)} 
 +   \varepsilon^0_{\mathrm g} 
 + 2 \varepsilon^0_{\mathrm u}
 - \frac{3}{4} U_{\mathrm g}
 - \frac{1}{2} U_{\mathrm u} 
 - \frac{5}{2} U' 
,
\\ \nonumber \\
E^{(3)}_{\mathrm{u}} &=& 
E^{(0)} 
 + 2 \varepsilon^0_{\mathrm g} 
 +   \varepsilon^0_{\mathrm u}
 - \frac{1}{2} U_{\mathrm g}
 - \frac{3}{4} U_{\mathrm u} 
 - \frac{5}{2} U'
,
\\ \nonumber \\
E^{(4)} 
&=&
E^{(0)} 
 + 2 \varepsilon^0_{\mathrm g} 
 + 2 \varepsilon^0_{\mathrm u}
 - \frac{1}{2} U_{\mathrm g}
 - \frac{1}{2} U_{\mathrm u} 
 - 2 U' 
.
\end{eqnarray}%
\label{eq:energy_1mol}%
\end{subequations}

Seven model parameters are to be determined, while
  ten energies are calculated for the respective 
configurations.
Thus, the expressions for 
the model parameters in terms of \textit{ab initio} energies
  are not expressed uniquely,  but we can obtain the same numerical 
 values irrespective of the expressions.
One possible way to relate 
 the configuration energies to the model parameters 
   is  given by
\begin{subequations}
\begin{eqnarray}
E_0
&=&
- \frac{5}{16} E^{(0)}
+ \frac{3}{8} E_{\mathrm{g}}^{(1)} 
+ \frac{3}{8} E_{\mathrm{u}}^{(1)} 
+ \frac{9}{16} E^{(4)} 
,\quad
\\ \nonumber \\
 \varepsilon^0_{\mathrm g} 
&=&
  \frac{1}{8} E^{(0)}
- \frac{1}{2} E_{\mathrm{g}}^{(1)} 
\nonumber \\ &&{}
+ \frac{3}{8} E_{\mathrm{g,s,1}}^{(2)} 
- \frac{3}{8} E_{\mathrm{g,s,2}}^{(2)} 
+ \frac{3}{8} E^{(4)} 
,
\\ \nonumber \\
  \varepsilon^0_{\mathrm{u}}
&=& 
  \frac{1}{8} E^{(0)}
- \frac{1}{2} E_{\mathrm{u}}^{(1)} 
\nonumber \\ &&{}
- \frac{3}{8} E_{\mathrm{g,s,1}}^{(2)} 
+ \frac{3}{8} E_{\mathrm{g,s,2}}^{(2)} 
+ \frac{3}{8} E^{(4)} 
,
\\ \nonumber \\
    U_{\mathrm g}
&=&
    E^{(0)}
- 2 E_{\mathrm{g}}^{(1)} 
+   E_{\mathrm{g,s,1}}^{(2)} 
,
\\ \nonumber \\
 U_{\mathrm u} 
&=&
    E^{(0)}
- 2 E_{\mathrm{u}}^{(1)} 
+   E_{\mathrm{g,s,2}}^{(2)} 
,
\\ \nonumber \\
  U'
&=&
  \frac{1}{4} E^{(0)} 
- \frac{1}{4} E_{\mathrm{g,s,1}}^{(2)} 
- \frac{1}{4} E_{\mathrm{g,s,2}}^{(2)} 
+ \frac{1}{4} E^{(4)} 
,
\\ \nonumber \\
 J_{\mathrm H} 
&=& 
  E^{(2)}_{\mathrm{u,s}} 
- E^{(2)}_{\mathrm{u,t}} 
.
\end{eqnarray}%
\end{subequations}
We note that $J_\mathrm{H}$ can also be derived from
  the off-diagonal pair-hopping term
 $2\langle \psi^{(2)}_{\mathrm{g,s,1}}| H 
  | \psi^{(2)}_{\mathrm{g,s,2}} \rangle $.

\subsection{Results}

\begin{table}[b]
\caption{
Estimated parameters for the (TTM-TTP)$^+$ and [Au(tmdt)$_2$] molecules.
All energies are in eV.
}
\label{eq:parameter-1mol}
\begin{ruledtabular}
\begin{tabular}{c|cc}
 &  TTM-TTP$^+$ & [Au(tmdt)$_2$] 
\\\hline 
Level energy $\varepsilon^0_{\mathrm g}$  
&  $-8.63$ & $-5.40$
\\
Level energy $\varepsilon^0_{\mathrm u}$ 
&  $-8.21$ & $-5.66$
\\
$U_{\mathrm{g}}$ 
&  $\phantom{+}3.70$ & $\phantom{+}3.49$
\\
$U_{\mathrm{u}}$
&  $\phantom{+}3.90$ & $\phantom{+}3.45$
\\
$U'$
&  $\phantom{+}2.82$ & $\phantom{+}2.48$
\\
$J_{\mathrm{H}}$
&  $\phantom{+}3.19$ & $\phantom{+}3.65$
\end{tabular} 
\end{ruledtabular}
\end{table}

The model parameters obtained in this way are summarized in 
 Table \ref{eq:parameter-1mol}.
The on-site Coulomb repulsions $U_\mathrm{g}$ and
 $U_\mathrm{u}$ are comparable to 
  the values for the BEDT-TTF molecule $\sim 4.2 $ eV.
  \cite{Scriven2009a,Scriven2009b}
We have also applied the present scheme to the TTF molecule and obtained 
  large value of the ``on-site'' Coulomb interaction 
   $\sim 6.2$ eV.
By comparing the values of $U_\mathrm{g}$ and $U_\mathrm{u}$
 for TTM-TTP$^+$,
we find a relatively
  smaller value for $U_\mathrm{g}$, which can be explained by 
  noting the left-right bonding character of the u MO while 
the g MO is anti-bonding,
as seen in Fig.\ \ref{fig:MO}.
On the other hand, the values of $U_\mathrm{g}$ and $U_\mathrm{u}$
for [Au(tmdt)$_2$] are almost degenerate as a consequence
of the enhanced delocalization onto a larger system.
The magnitudes
 of inter-orbital interactions, $U'$ and $J_\mathrm{H}$,
   are large compared with transition
  metal atom situations, satisfying
  $U_{\mathrm{atom}}=U_{\mathrm{atom}}'+J_{\mathrm{H\, atom}}$.
In contrast with the transition metal atoms displaying a centrosymmetric 
  potential, there is no constraint relation among the couplings, 
  $U_\mathrm{g}$, $U_\mathrm{u}$, $U'$, and $J_{\mathrm{H}}$.
Our results show that
  $(U_\mathrm{g}+U_\mathrm{u})/2\approx 3\sim 4$ eV and
  $(U'+J_\mathrm{H}) \approx 6 $ eV.
As a simplest and most intuitive example for the parameter evaluation
  of two-orbital system, we can consider the hydrogen molecule H$_2$
  where the full CI calculations can be performed
  and we can get an insight into the large Hund coupling 
  in molecular systems.\cite{SI}

\section{Inter-molecular interactions}

By extending the analysis of Sec.\ III,
   we evaluate the inter-molecular interactions 
by focusing on 
  two neighboring molecules, i.e., dimer, in the crystal.

\subsection{Construction of the two-orbital extended Hubbard Hamiltonian}

\begin{figure}[b]
\begin{center}
\includegraphics[width=8cm,bb=0 0 439 325]{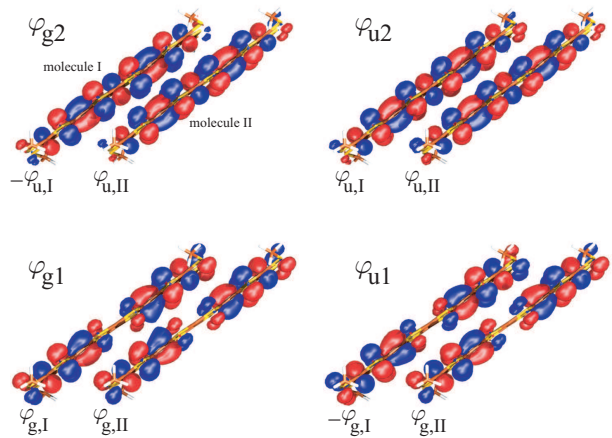}
\end{center}
\caption{
Side view of the ``effective molecular orbitals''
 for the two-molecule dimer system.
The MOs of each TTM-TTP molecule are the same as those calculated  
 for the isolated molecule.
The wavefunctions $\varphi_{\mathrm{g1}}$ and $\varphi_\mathrm{u1}$
 ($\varphi_{\mathrm{g2}}$ and $\varphi_\mathrm{u2}$)
 are given by  the linear combinations
 of two g (u) MOs obtained from the isolated-molecule calculations.
}
\label{fig:MO-2mol}
\end{figure}

A similar strategy can be applied to 
  obtain the expression of the two-orbital 
 Hamiltonian for the two-molecule dimer system.
The molecule index 
  is given by $\I$ and $\II$.
Each molecule has gerade and ungerade MOs 
 reflecting the symmetry with respect to the each inversion center.
 For the gerade and ungerade MOs on molecule $\I$, we adopt 
  the notations 
$\varphi_{\mathrm{g},\I}$ and $\varphi_{\mathrm{u},\I}$, respectively
($\varphi_{\mathrm{g},\II}$
 and  $\varphi_{\mathrm{u},\II}$ for molecule $\II$).
The MOs correspond to the strictly non-interacting situation. 
Practically,
they were determined by pulling apart 
the 
  two molecules.
The respective MOs are shown in Fig.\ \ref{fig:MO-2mol}.
The inversion transformations
   with respect to the inversion centers located on the molecule 
 ($\bm{R}_\I$ and $\bm{R}_\II$)
  are represented as, e.g.,
$\varphi_{\mathrm{g}, \I}(\bm R_\I+\bm r) 
  = \varphi_\mathrm{\mathrm{g}, \I}(\bm R_\I-\bm r)$ and 
$\varphi_\mathrm{\mathrm{u}, \I}(\bm R_\I +\bm r)
  = -\varphi_\mathrm{\mathrm{u}, \I}(\bm R_\I-\bm r)$.
From the x-ray structure analysis,
it has been shown that 
the system exhibits another inversion center located at midpoint  between 
   two molecules. \cite{Mori1994,Suzuki2003}
For the inversion transformation around 
the midpoint,
    the wavefunctions of  the two molecules follow the relations
\begin{align}
\varphi_{\mathrm{g}, \I}(\bm r) = \varphi_\mathrm{\mathrm{g}, \II}(-\bm r) ,
\quad
\varphi_\mathrm{\mathrm{u}, \I}(\bm r) = -\varphi_\mathrm{\mathrm{u}, \II}(-\bm r) .
\label{eq:symmetry-2mol} 
\end{align}
The
$\varphi_{\mathrm{g}, \I}$,
$\varphi_{\mathrm{u}, \I}$,
$\varphi_{\mathrm{g}, \II}$, and 
$\varphi_{\mathrm{u}, \II}$ MOs
are used to define the model space.

\begin{figure}[t]
\includegraphics[width=6cm,bb=0 0 127 74]{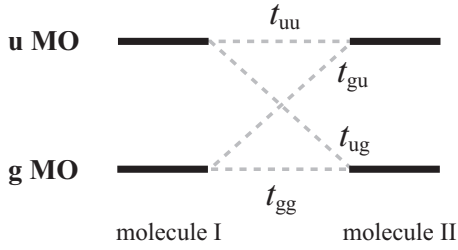}
\caption{
Tight-binding parameters for 
 the two-molecule dimer system.
We note $t_{\mathrm{gu}}=-t_{\mathrm{ug}}$
due to the presence of inversion center between two molecules.
}
\label{fig:model-gu}
\end{figure}

For the inter-molecular interactions, the most important parameters are
  transfer integrals between the molecules.
We define such transfer integrals
($\alpha,\beta=$ g or u) as:
\begin{eqnarray}
t_{\alpha\beta} &=&
- \langle \alpha_\I | H | \beta_\II \rangle
\nonumber \\
&=&
-\int d\bm{r} \, 
\varphi^*_{\alpha,\I}(\bm r) \, H  \, 
\varphi_{\beta,\II}(\bm r) .
\label{eq:transfer}
\end{eqnarray}
In the two-orbital system, four kinds of transfer integral are possible
 (Fig.\ \ref{fig:model-gu}),
  which are explicitly given by
\begin{subequations}
\begin{eqnarray}
t_{\mathrm{gg}} &=&
- \langle \mathrm g_\I | H | \mathrm g_\II \rangle,
\\ \nonumber \\
t_{\mathrm{uu}} &=& 
- \langle \mathrm u_\I | H | \mathrm u_\II \rangle,
\\ \nonumber \\
t_{\mathrm{gu}} &=&
- \langle \mathrm g_\I | H | \mathrm u_\II \rangle,
\\ \nonumber \\
t_{\mathrm{ug}}
&=&
-\langle \mathrm u_\I | H | \mathrm g_\II \rangle .
\end{eqnarray}%
\end{subequations}
The Hermitian condition indicates
  $\langle \alpha_\I | H | \beta_\II \rangle 
 = \langle \beta_\II | H | \alpha_\I \rangle$.
By applying the inversion transformation given by 
  Eq.\ (\ref{eq:symmetry-2mol}), 
 we find 
 $t_{\mathrm{gu}}=-t_{\mathrm{ug}}$.

In addition to the transfer integrals between the molecules,
   we have an extra one-electron interaction term
 in the two-molecule system.
If one picks up two molecules from the crystal, 
  the inversion center within each molecule is lost and 
 the mixtures of MOs, i.e., 
 $\langle \mathrm g_\I | H | \mathrm u_\I \rangle $ and 
$\langle \mathrm g_\II | H | \mathrm u_\II \rangle $
  become non-zero.
Such a symmetry-breaking interaction
is given by
\begin{eqnarray}
H_\Delta'
=
  \Delta\varepsilon_{\mathrm{gu}} \sum_\sigma
  ( c_{\mathrm{g},\I,\sigma}^\dagger c_{\mathrm{u},\I,\sigma}^{}
  - c_{\mathrm{g},\II,\sigma}^\dagger c_{\mathrm{u},\II,\sigma}^{}  + \mathrm{h.c.} )
.
\label{eq:H-Delta}
\end{eqnarray}
In terms of the \textit{ab initio} Hamiltonian $H$, 
 this coupling constant accounting for this symmetry-breaking interaction
  would be represented as
$\Delta \varepsilon_\mathrm{gu}=
  \frac{1}{2} \langle \mathrm g_\I | H | \mathrm u_\I \rangle 
- \frac{1}{2} \langle \mathrm g_\II | H | \mathrm u_\II \rangle $.
In the crystal, such interaction disappears 
due to the periodicity of the interactions.

In analogy with Eq.\ (\ref{eq:H_1mol-bare}),
 the electron-electron interaction terms are expressed as
\begin{eqnarray}
&& 
( \alpha_i \alpha'_{i'} , \beta_{j} \beta'_{j'} )
=
\iint d{\bm r}_1 d{\bm r}_2
\nonumber \\ && {} \quad \times 
\varphi_{\alpha,i}^* ({\bm r}_1) \varphi_{\alpha',i'}({\bm r}_1) 
\frac{1}{|{\bm r}_1-{\bm r}_2|}
\varphi_{\beta,j}^* ({\bm r}_2) \varphi_{\beta',j'}   ({\bm r}_2) ,
\nonumber \\
\label{eq:ee-integral}
\end{eqnarray}
where $\alpha,\beta=$ g or u, and the 
  molecule index is given by  $i,j,i',j'(=\I \mbox{ or } \II)$.  
We assume that all the wavefunctions are real and 
the relation of Eq.\ (\ref{eq:relation}) still holds in the present 
  case.
Under the inversion transformation ($\bm r\to -\bm r$)
given by Eq.\ (\ref{eq:symmetry-2mol}),
we find several relations, e.g.,
$ ( \mathrm{g}_\I \mathrm{u}_\I , \mathrm{u}_\I \mathrm{u}_\I )
=- ( \mathrm{g}_\II  \mathrm{u}_\II , \mathrm{u}_\II  \mathrm{u}_\II )$.
Since the MOs are localized on each molecule, 
  we find that the ``two-center'' integrals
  given by, e.g., 
$( \alpha_\I \alpha'_{\II} , \beta_{\II} \beta'_{\I} )$
and
$( \alpha_\I \alpha'_{\I} , \beta_{\I} \beta'_{\II} )$
are negligibly small due to 
small overlaps
   and cannot be determined within the numerical error.
The interactions expressed as
  $( \alpha_\I \alpha'_{\I}, \beta_{\I} \beta'_{\I} )$, 
  $( \alpha_\II \alpha'_{\II} , \beta_{\II} \beta'_{\II} )$,
  $( \alpha_\I \alpha'_{\I} , \beta_{\II} \beta'_{\II} )$,
  and
  $( \alpha_\II \alpha'_{\II}, \beta_{\I} \beta'_{\I} )$
  are the only relevant interactions.
We also note here that 
  the hopping parameters, Eq.\ (\ref{eq:transfer}), have been 
 defined in the one-electron picture and some of two-electron integrals, 
 e.g., $(\mathrm{g}_\I \mathrm{g}_\I, \mathrm{u}_\I \mathrm{u}_\II)$ and
 $(\mathrm{g}_\I \mathrm{u}_\I, \mathrm{u}_\I \mathrm{u}_\II)$, 
can contribute to them in the present two-orbital systems.
However, these contributions are small two-center integrals and thus
 the simple definition of the transfer integrals can be justified.
Depending on the choice of indices $\alpha, \alpha', \beta, \beta'$
  and $i,i',j,j'$, the interactions (\ref{eq:ee-integral})
can be classified into five categories:
(i) all the indices are identical, 
(ii) three-indices are identical, 
(iii) two pairs of identical indices,
(iv) all indices are different, and
(v) one pair of identical indices.
We here examine each case separately.

(i)
The interactions are nothing but the local on-site Coulomb interactions,
given by
$U_{\mathrm{g}} =( \mathrm g_\I \mathrm g_\I , \mathrm g_\I \mathrm g_\I )
= ( \mathrm g_\II \mathrm g_\II , \mathrm g_\II \mathrm g_\II )$
and 
$U_{\mathrm{u}} 
= ( \mathrm u_\I \mathrm u_\I , \mathrm u_\I \mathrm u_\I )
= (\mathrm u_\II \mathrm u_\II , \mathrm u_\II \mathrm u_\II )$,
which are the same relations as Eq.\ (\ref{eq:UgUu}).

(ii)
 The possible interactions are
$( \mathrm{g}_\I \mathrm{g}_\I , \mathrm{g}_\I \mathrm{u}_\I )$,
$( \mathrm{u}_\I \mathrm{u}_\I , \mathrm{u}_\I \mathrm{g}_\I )$,
  and the interactions with the molecular index $\I$ and $\II$ interchanged.
These couplings vanish 
due to the inversion transformation within each molecule.

(iii)
In addition to the intra-molecular interactions 
  $J_\mathrm{H}$ and $U'$, 
we introduce the couplings:
$V_\mathrm{gg} = ( \mathrm{g}_\I \mathrm{g}_\I , \mathrm{g}_\II \mathrm{g}_\II )$,
$V_\mathrm{uu} = ( \mathrm{u}_\I \mathrm{u}_\I , \mathrm{u}_\II \mathrm{u}_\II )$,
$V_\mathrm{gu} = ( \mathrm{g}_\I \mathrm{g}_\I,  \mathrm{u}_\II \mathrm{u}_\II )$, 
and
$V_\mathrm{ug} =( \mathrm{u}_\I \mathrm{u}_\I , \mathrm{g}_\II \mathrm{g}_\II )$,
where $V_{\mathrm{gu}}=V_{\mathrm{ug}}$.
These represent the ``inter-site'' Coulomb repulsions.

(iv)
In this category, 
the only relevant parameter is given by  
\begin{eqnarray}
I &=& 
( \mathrm{g}_\I \mathrm{u}_\I ,\mathrm{g}_\II \mathrm{u}_\II )
=( \mathrm{g}_\I \mathrm{u}_\I ,\mathrm{u}_\II \mathrm{g}_\II )
\nonumber \\ &=&
( \mathrm{u}_\I \mathrm{g}_\I ,\mathrm{u}_\II \mathrm{g}_\II )
=( \mathrm{u}_\I \mathrm{g}_\I ,\mathrm{g}_\II \mathrm{u}_\II )
.
\label{eq:I-2mol}
\end{eqnarray}
This term can be regarded as an ``orbital exchange'' between 
  neighboring molecules.

(v)
In this category, 
we have only two independent parameters, given by
\begin{subequations}
\begin{eqnarray}
X_\mathrm{g}
&=&
( \mathrm{g}_\I \mathrm{g}_\I , \mathrm{g}_\II \mathrm{u}_\II )
=
( \mathrm{g}_\I \mathrm{g}_\I , \mathrm{u}_\II \mathrm{g}_\II )
\nonumber \\
&=&
-
( \mathrm{g}_\II \mathrm{g}_\II , \mathrm{g}_\I \mathrm{u}_\I )
=
-
( \mathrm{g}_\II \mathrm{g}_\II , \mathrm{u}_\I \mathrm{g}_\I ),
\nonumber \\
\\ 
X_\mathrm{u}
&=&
( \mathrm{u}_\I \mathrm{u}_\I , \mathrm{g}_\II \mathrm{u}_\II )
=
( \mathrm{u}_\I \mathrm{u}_\I , \mathrm{u}_\II \mathrm{g}_\II )
\nonumber \\
&=&
-
( \mathrm{u}_\II \mathrm{u}_\II , \mathrm{g}_\I \mathrm{u}_\I )
=
-
( \mathrm{u}_\II \mathrm{u}_\II , \mathrm{u}_\I \mathrm{g}_\I ).
\end{eqnarray}
\end{subequations}

As in the case of the isolated molecule,
we introduce the density operators in the normal-ordered form:
$n_{\mathrm{g},i,\sigma}= (c_{\mathrm{g},i,\sigma}^\dagger c_{\mathrm{g},i,\sigma}^{}
 - \frac{3}{4})$ 
and
$n_{\mathrm{u},i,\sigma}= (c_{\mathrm{u},i,\sigma}^\dagger c_{\mathrm{u},i,\sigma}^{}
 - \frac{3}{4})$
where $i=\I,\II$.
Thus, 
in terms of model parameters, 
the full extended Hubbard Hamiltonian for the two-molecule system 
 is expressed as
\begin{eqnarray}
H_{2\mbox{-}\mathrm{mol}}
&=& 
\sum_{j=\I,\II} H_{1\mbox{-}\mathrm{mol}}
+ \sum_{j=\I,\II} 
  \bigl(  \Delta\varepsilon_{\mathrm{g}} n_{\mathrm{g},j}
        + \Delta\varepsilon_{\mathrm{u}} n_{\mathrm{u},j} \bigr)
\nonumber \\ && {}
-  t_{\mathrm{gg}} \sum_\sigma 
(c_{\mathrm{g},\I,\sigma}^\dagger c_{\mathrm{g},\II,\sigma}^{} +\mathrm{h.c.})
\nonumber \\ && {}
-  t_{\mathrm{uu}}  \sum_\sigma 
(c_{\mathrm{u},\I,\sigma}^\dagger c_{\mathrm{u},\II,\sigma}^{} +\mathrm{h.c.})
\nonumber \\ && {}
-  t_{\mathrm{gu}}  \sum_\sigma 
(c_{\mathrm{g},\I,\sigma}^\dagger c_{\mathrm{u},\II,\sigma}^{} 
-c_{\mathrm{u},\I,\sigma}^\dagger c_{\mathrm{g},\II,\sigma}^{} +\mathrm{h.c.})
\nonumber \\
&& {}
+ 
       V_{\mathrm{gg}} n_{\mathrm{g},\I} n_{\mathrm{g},\II}
 +     V_{\mathrm{uu}} n_{\mathrm{u},\I} n_{\mathrm{u},\II}
\nonumber \\ && {}
 +     V_{\mathrm{gu}} \left( n_{\mathrm{g},\I} n_{\mathrm{u},\II}
                         + n_{\mathrm{u},\I} n_{\mathrm{g},\II} \right)
\phantom{\frac{Z}{Z}}
\nonumber\\
&&
+  
I \smash[b]{ \sum_{\sigma,\sigma'} } \left(
   c_{\mathrm{g},\I,\sigma}^\dagger c_{\mathrm{u},\I,\sigma}
   c_{\mathrm{g},\II,\sigma'}^\dagger c_{\mathrm{u},\II,\sigma'}
\right. \nonumber \\ && {} \left. \qquad {}
 + c_{\mathrm{g},\I,\sigma}^\dagger c_{\mathrm{u},\I,\sigma}
   c_{\mathrm{u},\II,\sigma'}^\dagger c_{\mathrm{g},\II,\sigma'} 
 + \mathrm{h.c.}
\right)
\nonumber\\
&& {}
+     
X_{\mathrm{g}} \sum_\sigma
\left[
 n_{\mathrm{g},\I}
( c_{\mathrm{g},\II,\sigma}^\dagger c_{\mathrm{u},\II,\sigma} + \mathrm{h.c.} )
- (\I \leftrightarrow \II)
\right]
\nonumber \\ && {}
+
X_{\mathrm{u}} \sum_\sigma
\left[
 n_{\mathrm{u},\I}
( c_{\mathrm{g},\II,\sigma}^\dagger c_{\mathrm{u},\II,\sigma} + \mathrm{h.c.} )
- (\I \leftrightarrow \II)
\right]
\nonumber \\ && {}
+H_\Delta'
+\Delta E_{0} 
,
\label{eq:H-2mol}
\end{eqnarray}
where $H_{\mathrm{1\mbox{-}mol}}$ is the 
  Hamiltonian for the isolated molecule given in Eq.\
  (\ref{eq:singlemolecule}).
The $\Delta \varepsilon_\mathrm{g}$ and
  $\Delta \varepsilon_\mathrm{u}$ terms represent 
   the energy-level shift due to the neighboring molecules, 
  i.e., so-called crystal-field effect.
These terms also include the Hartree contributions arising from 
 the intermolecular density-density interactions, 
$V_\mathrm{gg}$, $V_{\mathrm{uu}}$, and $V_\mathrm{gu}$,
as in the isolated molecule [Eq.\ (\ref{eq:epsilon0})].
The Hartree contributions from the $X_\mathrm{g}$ 
 and $X_\mathrm{u}$ interactions can be included into the $H_\Delta'$ term 
[Eq.\ (\ref{eq:H-Delta})].
The term $\Delta E_{0}$ represents the constant energy shift.

\subsection{Possible configurations}

For the two-molecule systems, 
 four MOs must be considered.
Due to the presence of the transfer integrals between the molecules, 
  the wavefunctions for the two-molecule system
    are given by the linear combinations of MOs for the isolated 
  molecules.
The fragment MOs which exhibit g and u characters within each molecule,
namely ($\varphi_{\mathrm{g,\I}}, \varphi_{\mathrm{u,\I}}$) and
 ($\varphi_{\mathrm{g,\II}}, \varphi_{\mathrm{u,\II}}$), 
  were used to build the symmetry-adapted MOs,  given by
\begin{subequations}
\begin{eqnarray}
\varphi_{\mathrm{g}1}&=&
 \frac{1}{\sqrt{2}}   (\varphi_{\mathrm{g},\I}+\varphi_{\mathrm{g},\II}),
\\  \nonumber \\
\varphi_{\mathrm{u}1}&=&
 \frac{1}{\sqrt{2}}    (-\varphi_{\mathrm{g},\I}+\varphi_{\mathrm{g},\II}),
\\ \nonumber \\
\varphi_{\mathrm{g}2}&=&
 \frac{1}{\sqrt{2}}    (-\varphi_{\mathrm{u},\I}+\varphi_{\mathrm{u},\II}),
\\ \nonumber \\
\varphi_{\mathrm{u}2}&=&
 \frac{1}{\sqrt{2}}    (\varphi_{\mathrm{u},\I}+\varphi_{\mathrm{u},\II}),
\end{eqnarray}%
\label{eq:MO-2mol}%
\end{subequations}
  and  drawn in Fig.\ \ref{fig:MO-2mol}.

As in the case of the isolated molecule system, 
different configurations were considered.
In order to specify each configuration on the new MO basis
[Eq.\ (\ref{eq:MO-2mol})],
 we introduce the following:
\begin{eqnarray}
\left| 
      \begin{array}{cc} 
\displaystyle
         \frac{\mbox{\scriptsize g$_2$}}{\phantom{-}} & 
\displaystyle
         \frac{\mbox{\scriptsize u$_2$}}{\phantom{-}} \\
\displaystyle
         \frac{\mbox{\scriptsize g$_1$}}{\phantom{-}} & 
\displaystyle
         \frac{\mbox{\scriptsize u$_1$}}{\phantom{-}} \\
      \end{array}
\right\rangle
.
\end{eqnarray}
In the two-molecule systems [(TTM-TTP)$^-$]$_2$ and
 [Au(tmdt)$_2$]$_2$, 
six electrons are likely to occupy these four MOs.
In order to estimate the magnitudes of the Coulomb interactions,
 we have to add/remove electrons from the six-electron
 ground state. 
In the present analysis, we consider all the configurations 
  with $n=0$, $1$, and $2$ electrons.
The extensions to the states with $n\ge 3$ are straightforward. 
However, 
we will see that the information is sufficient to 
  evaluate the model parameters.

The zero-electron state is given by
\begin{eqnarray}
|\psi^{(0)}\rangle &=&
    \left| 
      \begin{array}{cc}
           -  &  -  \\
           -  &  -
      \end{array}
    \right\rangle
= | 0\rangle ,
\end{eqnarray}
while the one-electron states are
\begin{subequations}
\begin{eqnarray}
|\psi_{\mathrm{g,1}}^{(1)} \rangle &=&
    \left| 
      \begin{array}{cc}
           -            &  -  \\
         \, \uparrow \, &  -
      \end{array}
    \right\rangle
=
c_{\mathrm{g_1},\uparrow}^\dagger |0\rangle
,
\\ \nonumber \\
| \psi_{\mathrm{g,2}}^{(1)} \rangle &=&
    \left| 
      \begin{array}{cc}
         \, \uparrow \, &  -  \\
           -            &  -
      \end{array}
    \right\rangle
=
c_{\mathrm{g_2},\uparrow}^\dagger |0\rangle
,
\\ \nonumber \\
| \psi_{\mathrm{u,1}}^{(1)} \rangle &=&
    \left| 
      \begin{array}{cc}
         - &  -            \\
         - & \, \uparrow \,
      \end{array}
    \right\rangle
=
c_{\mathrm{u_1},\uparrow}^\dagger |0\rangle
,
\\ \nonumber \\
| \psi_{\mathrm{u,2}}^{(1)} \rangle &=&
    \left| 
      \begin{array}{cc}
         - &  \, \uparrow \, \\
         - &  -
      \end{array}
    \right\rangle
=
c_{\mathrm{u_2},\uparrow}^\dagger |0\rangle
,
\end{eqnarray}%
\label{eq:config-2mol-1el}%
\end{subequations}
where $c_{\mathrm{g_1},\sigma}$
$c_{\mathrm{u_1},\sigma}$, 
$c_{\mathrm{g_2},\sigma}$, and 
$c_{\mathrm{u_2},\sigma}$
  represent the 
   annihilation operators corresponding to 
 the wavefunctions,
   $\varphi_{\mathrm{g}1}$,
   $\varphi_{\mathrm{u}1}$,
   $\varphi_{\mathrm{g}2}$, and
   $\varphi_{\mathrm{u}2}$, respectively, 
 given in Fig.\ \ref{fig:MO-2mol} and Eq.\ (\ref{eq:MO-2mol}).

\begin{table*}
\caption{
Possible configurations for the two-electron states 
in the two-molecule dimer systems,
the corresponding expressions in the molecular index $\I$ and $\II$, and
the energy expectation values
represented in terms of 
the parameters in the extended Hubbard Hamiltonian
$E_{\nu}^{(2)}=\langle \psi_{\nu}^{(2)} |  H_{\mathrm{2\mbox{-}mol}}
               | \psi_{\nu}^{(2)} \rangle$, where
$ E^{(2)}
=
E^{(0)}
+ \varepsilon_{\mathrm{g}} 
+ \varepsilon_{\mathrm{u}}
- 3 (U_{\mathrm{g}}+U_{\mathrm{u}}) /4
- 5 U' /2
- 3 (V_{\mathrm{gg}}+V_{\mathrm{uu}})/2
- 5 V_{\mathrm{gu}} /2 $.
The suffix s and t represent the spin-singlet and spin-triplet 
  states, respectively.
}
\label{table:2mol2el}
\begin{ruledtabular}
{ \renewcommand\arraystretch{1.5}
\begin{tabular}{llll}
&Configurations 
& Representation on the molecular-based fragment MOs  
& Energy expectation values
\\ \hline
\multirow{6}{*}{Gerade, singlet}
&
$|\psi_{\mathrm{g,s,1}}^{(2)}\rangle
=
    \left| 
      \begin{array}{cc}
           -            &  -  \\
         \uparrow\downarrow &  -
      \end{array}
    \right\rangle
$
&
$\begin{array}{l}
\frac{1}{2}
\left(
  |\,\mathrm{g_\I\bar{g}_\I}\rangle
+ |\,\mathrm{g_\II\bar{g}_\II}\rangle
+ |\,\mathrm{g_\I\bar{g}_\II}\rangle
+ |\,\mathrm{g_\II\bar{g}_\I}\rangle
\right)
\end{array}$
&
$\begin{array}{l}
E^{(2)}
+   \varepsilon_{\mathrm{g}}
-   \varepsilon_{\mathrm{u}}
- \frac{1}{4} U_{\mathrm{g}}
+ \frac{3}{4} U_{\mathrm{u}} 
- \frac{1}{2} U' 
\\ \qquad
-             V_{\mathrm{gg}}
+ \frac{3}{2} V_{\mathrm{uu}}
- \frac{1}{2} V_{\mathrm{gu}} 
- 2 t_{\mathrm{gg}} 
\end{array}$
\\ 
&
$|\psi^{(2)}_{\mathrm{g,s,2}}\rangle
=
    \left| 
      \begin{array}{cc}
          \, \downarrow \,   & - \\
          \, \uparrow \,  &  -
      \end{array}
    \right\rangle
$
&
$\begin{array}{l}
-
\frac{1}{2\sqrt{2}}
(
 |\,\mathrm{g_\I \bar u_\I}\rangle + |\,\mathrm{u_\I \bar g_\I}\rangle
-|\,\mathrm{g_\II \bar u_\II}\rangle - |\,\mathrm{u_\II \bar g_\II}\rangle
)
\\ \qquad
+
 \frac{1}{2\sqrt{2}}
(
 |\,\mathrm{g_\I \bar u_\II}\rangle + |\,\mathrm{u_\II \bar g_\I}\rangle
-|\,\mathrm{g_\II \bar u_\I}\rangle - |\,\mathrm{u_\I \bar g_\II}\rangle
)
\end{array}$
&
$\begin{array}{l}
E^{(2)}
+ \frac{3}{8} J_{\mathrm{H}}
- \frac{1}{2} I  
+ \left( t_\mathrm{uu} - t_\mathrm{gg} \right)
\end{array}$
\\ 
&
$|\psi_{\mathrm{g,s,3}}^{(2)}\rangle
=
    \left| 
      \begin{array}{cc}
         \uparrow\downarrow &  -  \\
         -   &  -
      \end{array}
    \right\rangle
$
&
$\begin{array}{l}
\frac{1}{2}
\left(
  |\,\mathrm{u_\I\bar{u}_\I}\rangle
+ |\,\mathrm{u_\II\bar{u}_\II}\rangle
- |\,\mathrm{u_\I\bar{u}_\II}\rangle
- |\,\mathrm{u_\II\bar{u}_\I}\rangle
\right) 
\end{array}$
&
$
\begin{array}{l}
E^{(2)}
-   \varepsilon_{\mathrm{g}} 
+   \varepsilon_{\mathrm{u}}
+ \frac{3}{4} U_{\mathrm{g}} 
- \frac{1}{4} U_{\mathrm{u}} 
- \frac{1}{2} U' 
\\ \qquad
+ \frac{3}{2} V_{\mathrm{gg}}
-             V_{\mathrm{uu}}
- \frac{1}{2} V_{\mathrm{gu}}
+ 2 t_{\mathrm{uu}} 
\end{array}
$
\\ 
&
$| \psi_{\mathrm{g,s,4}}^{(2)} \rangle
=
    \left| 
      \begin{array}{cc}
           -            &  -  \\
           -  & \uparrow\downarrow 
      \end{array}
    \right\rangle
$
&
$\begin{array}{l}
\frac{1}{2}
\left(
  |\,\mathrm{g_\I\bar{g}_\I}\rangle
+ |\,\mathrm{g_\II\bar{g}_\II}\rangle
- |\,\mathrm{g_\I\bar{g}_\II}\rangle
- |\,\mathrm{g_\II\bar{g}_\I}\rangle
\right)
\end{array}$
&
$
\begin{array}{l}
E^{(2)}
+   \varepsilon_{\mathrm{g}} 
-   \varepsilon_{\mathrm{u}}
- \frac{1}{4} U_{\mathrm{g}} 
+ \frac{3}{4} U_{\mathrm{u}} 
- \frac{1}{2} U' 
\\ \qquad
-             V_{\mathrm{gg}}
+ \frac{3}{2} V_{\mathrm{uu}}
- \frac{1}{2} V_{\mathrm{gu}} 
+ 2 t_{\mathrm{gg}} 
\end{array}
$
\\ 
&
$| \psi^{(2)}_{\mathrm{g,s,5}} \rangle
=
    \left| 
      \begin{array}{cc}
          - & \, \downarrow \, \\
          - & \, \uparrow \,  
      \end{array}
    \right\rangle
$
&
$
\begin{array}{l}
-
\frac{1}{2\sqrt{2}}
(
 |\,\mathrm{g_\I \bar u_\I}\rangle + |\,\mathrm{u_\I \bar g_\I}\rangle
-|\,\mathrm{g_\II \bar u_\II}\rangle - |\,\mathrm{u_\II \bar g_\II}\rangle
)
\\ \qquad
-
 \frac{1}{2\sqrt{2}}
(
 |\,\mathrm{g_\I \bar u_\II}\rangle + |\,\mathrm{u_\II \bar g_\I}\rangle
-|\,\mathrm{g_\II \bar u_\I}\rangle - |\,\mathrm{u_\I \bar g_\II}\rangle
)
\end{array}
$
&
$\begin{array}{l}
E^{(2)}
+ \frac{3}{8} J_{\mathrm{H}}
- \frac{1}{2} I  
- \left(  t_\mathrm{uu}- t_\mathrm{gg} \right)
\end{array}$
\\ 
&
$| \psi_{\mathrm{g,s,6}}^{(2)} \rangle
=
    \left| 
      \begin{array}{cc}
         - & \uparrow\downarrow  \\
         -   &  -
      \end{array}
    \right\rangle
$
&
$\begin{array}{l}
\frac{1}{2}
\left(
  |\,\mathrm{u_\I\bar{u}_\I}\rangle
+ |\,\mathrm{u_\II\bar{u}_\II}\rangle
+ |\,\mathrm{u_\I\bar{u}_\II}\rangle
+ |\,\mathrm{u_\II\bar{u}_\I}\rangle
\right) 
\end{array}$
&
$
\begin{array}{l}
E^{(2)}
-   \varepsilon_{\mathrm{g}} 
+   \varepsilon_{\mathrm{u}}
+ \frac{3}{4} U_{\mathrm{g}} 
- \frac{1}{4} U_{\mathrm{u}} 
- \frac{1}{2} U' 
\\ \qquad
+ \frac{3}{2} V_{\mathrm{gg}}
-             V_{\mathrm{uu}}
- \frac{1}{2} V_{\mathrm{gu}}
- 2 t_{\mathrm{uu}} 
\end{array}
$
\\ \hline 
\multirow{2}{*}{Gerade, triplet}
&
$| \psi^{(2)}_{\mathrm{g,t,1}} \rangle
=
    \left| 
      \begin{array}{cc}
          \, \uparrow \,   & - \\
          \, \uparrow \,  &  -
      \end{array}
    \right\rangle
$
&
$\begin{array}{l}
-
\frac{1}{2}
(
 |\,\mathrm{g_\I u_\I}\rangle - |\,\mathrm{g_\II u_\II}\rangle
-|\,\mathrm{g_\I u_\II}\rangle + |\,\mathrm{g_\II u_\I}\rangle
)
\end{array}$
&
$\begin{array}{l}
E^{(2)}
- \frac{1}{8} J_{\mathrm{H}}
+ \frac{1}{2} I
+ (t_\mathrm{uu} - t_\mathrm{gg}) 
\end{array}$
\\ 
&
$| \psi^{(2)}_{\mathrm{g,t,2}} \rangle
=
    \left| 
      \begin{array}{cc}
          - & \, \uparrow \,  \\
          - & \, \uparrow \, 
      \end{array}
    \right\rangle
$
&
$\begin{array}{l}
-
\frac{1}{2}
(
 |\,\mathrm{g_\I u_\I}\rangle - |\,\mathrm{g_\II u_\II}\rangle
+|\,\mathrm{g_\I u_\II}\rangle - |\,\mathrm{g_\II u_\I}\rangle
)
\end{array}$
&
$\begin{array}{l}
E^{(2)}
- \frac{1}{8} J_{\mathrm{H}}
+ \frac{1}{2} I
- (t_\mathrm{uu} - t_\mathrm{gg}) 
\end{array}$
\\ \hline
\multirow{4}{*}{Ungerade, singlet}
&
$| \psi_{\mathrm{u,s,1}}^{(2)} \rangle
=
    \left|
      \begin{array}{cc}
           -            &  -  \\
        \, \uparrow \, &  \, \downarrow \,
      \end{array}
    \right\rangle$
&
$\begin{array}{l}
-\frac{1}{\sqrt{2}}
\left(
  |\,\mathrm{g_\I\bar{g}_\I}\rangle
- |\,\mathrm{g_\II\bar{g}_\II}\rangle
\right)
\end{array}$
&
$
\begin{array}{l}
E^{(2)}
+   \varepsilon_{\mathrm{g}} 
-   \varepsilon_{\mathrm{u}}
+ \frac{1}{4} U_{\mathrm{g}} 
+ \frac{3}{4} U_{\mathrm{u}} 
- \frac{1}{2} U' 
\\ \qquad
- \frac{3}{2} V_{\mathrm{gg}}
+ \frac{3}{2} V_{\mathrm{uu}}
- \frac{1}{2} V_{\mathrm{gu}} 
\end{array}
$
\\ 
&
$| \psi^{(2)}_{\mathrm{u,s,2}} \rangle
=
    \left| 
      \begin{array}{cc}
           -            &  \, \downarrow \,  \\
        \, \uparrow \, &  -
      \end{array}
    \right\rangle$
&
$
\begin{array}{l}
\frac{1}{2\sqrt{2}}
(
 |\,\mathrm{g_\I \bar u_\I}\rangle + |\,\mathrm{u_\I \bar g_\I}\rangle
+|\,\mathrm{g_\II \bar u_\II}\rangle + |\,\mathrm{u_\II \bar g_\II}\rangle
)
\\ \qquad
+ \frac{1}{2\sqrt{2}}
(
 |\,\mathrm{g_\I \bar u_\II}\rangle + |\,\mathrm{u_\II \bar g_\I}\rangle
+|\,\mathrm{g_\II \bar u_\I}\rangle + |\,\mathrm{u_\I \bar g_\II}\rangle
)
\end{array}
$
&
$\begin{array}{l}
E^{(2)}
+ \frac{3}{8} J_{\mathrm{H}}
+ \frac{1}{2} I 
- \left(  t_\mathrm{uu} +  t_\mathrm{gg} \right)
\end{array}$
\\ 
&
$| \psi^{(2)}_{\mathrm{u,s,3}} \rangle
=
    \left| 
      \begin{array}{cc}
        \, \uparrow \, &  -  \\
        - & \, \downarrow \,
      \end{array}
    \right\rangle$
&
$
\begin{array}{l}
\frac{1}{2\sqrt{2}}
(
 |\,\mathrm{g_\I \bar u_\I}\rangle + |\,\mathrm{u_\I \bar g_\I}\rangle
+|\,\mathrm{g_\II \bar u_\II}\rangle + |\,\mathrm{u_\II \bar g_\II}\rangle
)
\\ \qquad
- \frac{1}{2\sqrt{2}}
(
 |\,\mathrm{g_\I \bar u_\II}\rangle + |\,\mathrm{u_\II \bar g_\I}\rangle
+|\,\mathrm{g_\II \bar u_\I}\rangle + |\,\mathrm{u_\I \bar g_\II}\rangle
)
\end{array}
$
&
$\begin{array}{l}
E^{(2)}
+ \frac{3}{8} J_{\mathrm{H}}
+ \frac{1}{2} I 
+ \left(  t_\mathrm{uu} + t_\mathrm{gg} \right)
\end{array}$
\\ 
&
$| \psi_{\mathrm{u,s,4}}^{(2)} \rangle
=
    \left| 
      \begin{array}{cc}
          \, \uparrow \,   &  \, \downarrow \,  \\
           -  &  -
      \end{array}
    \right\rangle$
&
$\begin{array}{l}
-
\frac{1}{\sqrt{2}}
\left(
  |\,\mathrm{u_\I\bar{u}_\I}\rangle
- |\,\mathrm{u_\II\bar{u}_\II}\rangle
\right)
\end{array}$
&
$
\begin{array}{l}
E^{(2)}
-   \varepsilon_{\mathrm{g}} 
+   \varepsilon_{\mathrm{u}}
+ \frac{3}{4} U_{\mathrm{g}} 
+ \frac{1}{4} U_{\mathrm{u}} 
\\ \qquad
- \frac{1}{2} U' 
+ \frac{3}{2} V_{\mathrm{gg}}
- \frac{3}{2} V_{\mathrm{uu}}
- \frac{1}{2} V_{\mathrm{gu}} 
\end{array}
$
\\ \hline 
\multirow{4}{*}{Ungerade, triplet}
&
$| \psi_{\mathrm{u,t,1}}^{(2)} \rangle
=
    \left| 
      \begin{array}{cc}
           -            &  -  \\
        \, \uparrow \, &  \, \uparrow \,
      \end{array}
    \right\rangle$
&
$\begin{array}{l}
|\,\mathrm{g_\I g_\II}\rangle
\end{array}$
&
$
\begin{array}{l}
E^{(2)}
+   \varepsilon_{\mathrm{g}} 
-   \varepsilon_{\mathrm{u}}
- \frac{3}{4} U_{\mathrm{g}} 
+ \frac{3}{4} U_{\mathrm{u}} 
\\ \qquad
- \frac{1}{2} U' 
- \frac{1}{2} V_{\mathrm{gg}}
+ \frac{3}{2} V_{\mathrm{uu}}
- \frac{1}{2} V_{\mathrm{gu}} 
\end{array}
$
\\ 
&
$| \psi_{\mathrm{u,t,2}}^{(2)} \rangle
=
    \left| 
      \begin{array}{cc}
           -            &  \, \uparrow \,  \\
        \, \uparrow \, &  -
      \end{array}
    \right\rangle$
&
$\begin{array}{l}
\frac{1}{2} 
(
 |\,\mathrm{g_\I u_\I}\rangle
+|\,\mathrm{g_\II u_\II}\rangle
+|\,\mathrm{g_\I u_\II}\rangle
-|\,\mathrm{u_\I g_\II}\rangle
)
\end{array}$
&
$\begin{array}{l}
E^{(2)}
- \frac{1}{8} J_{\mathrm{H}}
- \frac{1}{2} I 
- ( t_\mathrm{uu} + t_\mathrm{gg} )
\end{array}$
\\ 
&
$| \psi_{\mathrm{u,t,3}}^{(2)} \rangle
=
    \left| 
      \begin{array}{cc}
        \, \uparrow \, &  -  \\
        - & \, \uparrow \,
      \end{array}
    \right\rangle$
&
$\begin{array}{l}
\frac{1}{2} 
(
-|\,\mathrm{g_\I u_\I}\rangle
-|\,\mathrm{g_\II u_\II}\rangle
+|\,\mathrm{g_\I u_\II}\rangle
-|\,\mathrm{u_\I g_\II}\rangle
)
\end{array}$
&
$\begin{array}{l}
E^{(2)}
- \frac{1}{8} J_{\mathrm{H}}
- \frac{1}{2} I 
+ ( t_\mathrm{uu} + t_\mathrm{gg} )
\end{array}$
\\ 
&
$| \psi_{\mathrm{u,t,4}}^{(2)} \rangle
=
    \left| 
      \begin{array}{cc}
          \, \uparrow \,   &  \, \uparrow \,  \\
           -  &  -
      \end{array}
    \right\rangle$
&
$\begin{array}{l}
-
 |\,\mathrm{u_\I u_\II}\rangle
\end{array}$
&
$
\begin{array}{l}
E^{(2)}
-   \varepsilon_{\mathrm{g}} 
+   \varepsilon_{\mathrm{u}}
+ \frac{3}{4} U_{\mathrm{g}} 
- \frac{3}{4} U_{\mathrm{u}} 
- \frac{1}{2} U' 
\\ \qquad
+ \frac{3}{2} V_{\mathrm{gg}}
- \frac{1}{2} V_{\mathrm{uu}}
- \frac{1}{2} V_{\mathrm{gu}} 
\end{array}
$
\end{tabular}
}
\end{ruledtabular}
\end{table*}

\begin{figure}[t]
\includegraphics[width=8cm,bb=0 0 337 321]{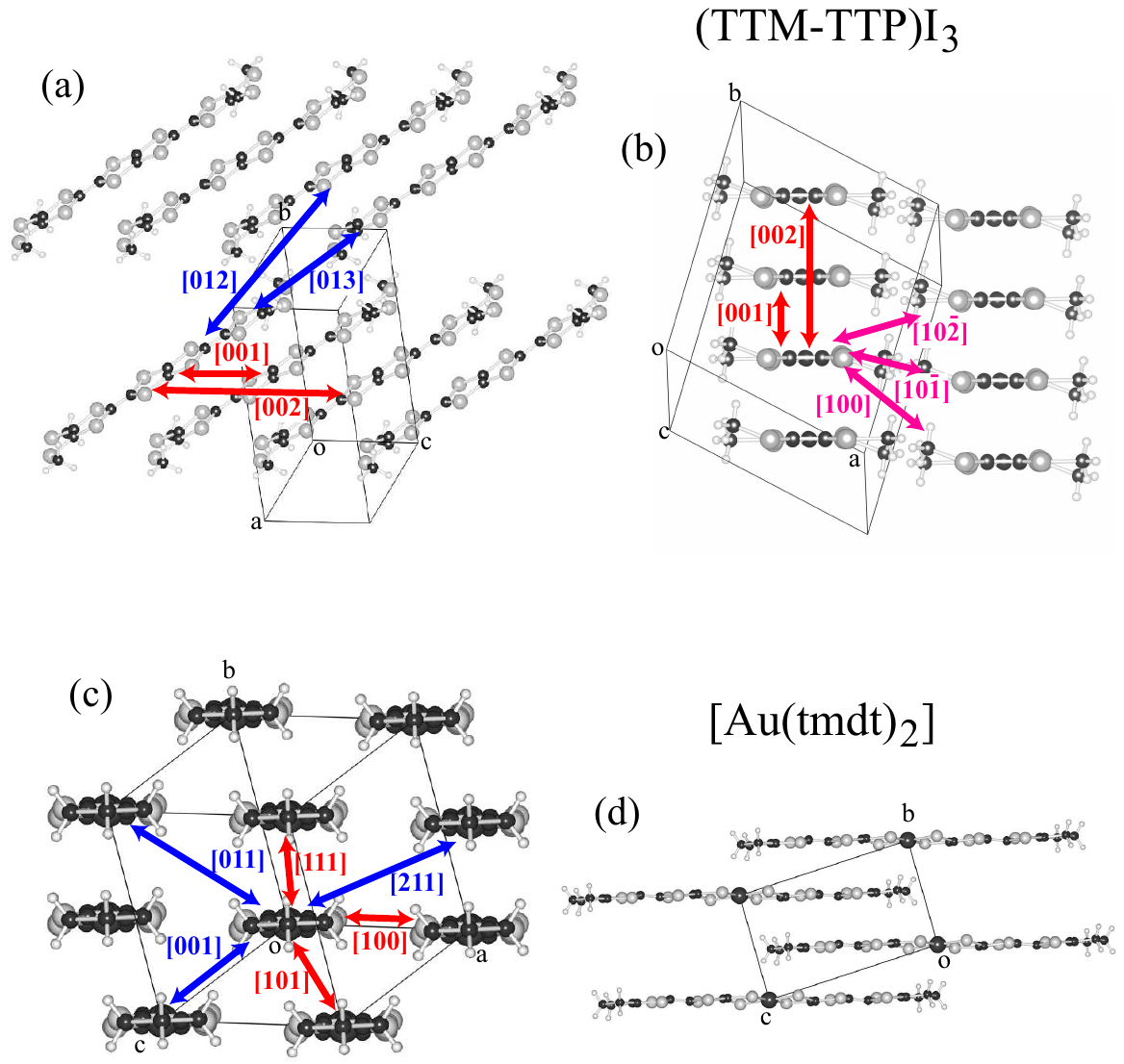}
\caption{
Crystal structure of (TTM-TTP)I$_3$ [(a) and
  (b)], and 
crystal structure of [Au(tmdt)$_2$] [(c) and (d)].
The target pairs of two neighboring molecules are shown by the arrows,
e.g.,
 [001] and [002] in (TTM-TTP)I$_3$ represent the nearest-neighboring and
  next-nearest neighboring molecule pairs 
  along the stacking $c$ direction.
}
\label{fig:crystal}
\end{figure}

We also consider all the two-electron configurations.
All the resulting states are listed in Table \ref{table:2mol2el},
depending on the g/u and singlet/triplet natures.

\subsection{\textit{Ab initio} calculations for the two-molecule dimer systems}
\label{abinitio-2mol}

The molecule pairs which we 
consider are 
extracted from the x-ray data, and 
are shown in 
  Fig.\ \ref{fig:crystal}.
In the case of [001] two-TTM-TTP-molecule pair,
   the ``molecular orbitals'' obtained from the ROHF calculation 
 are given in Fig.\ \ref{fig:MO-2mol}.
The assumption that we made here is that the MOs obtained in the 
  isolated-molecule calculation are similar to the MOs in crystal.

MR-CI calculations were performed to  
 extract full parameters by taking advantage of the
 wavefunction information.
In this scheme, 
one can access the off-diagonal elements of 
 the 
 \textit{ab initio} Hamiltonian from the knowledge of
  the eigenvalues and wavefunctions.
Such approach was originally developed for 
the situation with
 larger CI expansions in which effective parameters 
   can be extracted by projecting those onto a model space.
\cite{Bloch1958,Calzado2002a,Calzado2002b}

The MR-CI wavefunctions can be written as
\begin{eqnarray}
| \Psi_j \rangle  = 
\sum_{i} d_{ij} | \psi_i \rangle
,
\label{eq:CI}
\end{eqnarray}
  where the $|\psi_i\rangle $s
   are the configurations defined in Sec. IV B.
From the MR-CI calculations, 
  the energy eigenvalues $\{E_j\}$ and 
  eigenfunctions $\{d_{ij}\}$ were obtained 
 not only for the ground state but all 
  the excited states
contained in the model space.
The Schr\"odinger equation can be
  formally expressed as
\begin{eqnarray}
H 
\left(
\begin{array}{c}
d_{1j} \\ d_{2j} \\ \vdots \\ d_{Nj}
\end{array}
\right)
=
E_j 
\left(
\begin{array}{c}
d_{1j} \\ d_{2j} \\ \vdots \\ d_{Nj}
\end{array}
\right)
,
\end{eqnarray}
where $j=1,\ldots, N$, with $N$ being the number of possible states.
From  $\{d_{ij}\}$, 
we define the  unitary (orthogonal) matrix: 
\begin{eqnarray}
U= 
\left(
\begin{array}{cccc}
d_{11} & d_{12} & \ldots & d_{1N} \\ 
d_{21} & d_{22} &        & d_{2N} \\
\vdots &        & \ddots & \vdots \\
d_{N1} &        &        & d_{NN}
\end{array}
\right)
.
\end{eqnarray}
Together with the eigenvalues $\{E_j\}$,
the Hamiltonian can be written as
\begin{eqnarray}
H
= 
U
\left(
\begin{array}{cccc}
 E_1   &        &        &  \\ 
       & E_2    &        &  \\
       &        & \ddots &  \\
       &        &        &  E_N
\end{array}
\right)
U^{-1}
.
\label{eq:Habinitio}
\end{eqnarray}
Thus we can get not only the energy of each configuration, but also 
  the off-diagonal matrix elements of the Hamiltonian.

\subsection{Parameters mapping}

Evidently, the full Hamiltonian is expressed on the basis 
 $(\mathrm{g}_1,\mathrm{u}_1,\mathrm{g}_{2},\mathrm{u}_{2})$
whereas the model Hamiltonian
 is written on the molecular-based fragment MOs, 
 $(\mathrm{g}_\I,\mathrm{u}_{\I},\mathrm{g}_{\II},\mathrm{u}_{\II})$.
As can be seen from Fig.\ \ref{fig:MO-2mol} and Eq.\ (\ref{eq:MO-2mol}),
the correspondence between the basis
results in
\begin{subequations}
\begin{eqnarray}
c_{\mathrm{g_1},\sigma}
&=& \frac{1}{\sqrt{2}} 
(c_{\mathrm{g},\I,\sigma} + c_{\mathrm{g},\II,\sigma})
,
\\ \nonumber \\
c_{\mathrm{u_1},\sigma}
&=&
\frac{1}{\sqrt{2}} 
(-c_{\mathrm{g},\I,\sigma} + c_{\mathrm{g},\II,\sigma})
,
\\ \nonumber \\
c_{\mathrm{g_2},\sigma}
&=& 
\frac{1}{\sqrt{2}} 
(-c_{\mathrm{u},\I,\sigma} + c_{\mathrm{u},\II,\sigma})
,
\\ \nonumber \\
c_{\mathrm{u_2},\sigma}
&=& \frac{1}{\sqrt{2}} (c_{\mathrm{u},\I,\sigma} + c_{\mathrm{u},\II,\sigma})
.
\end{eqnarray}%
\label{eq:correspondence}%
\end{subequations}
By using these relations, 
  we rewrite all the configurations 
 into the molecular-based fragment MO basis, 
and derive the expression 
of Eq.\ (\ref{eq:Habinitio})
  in terms of the parameters of the model Hamiltonian 
  [Eq.\ (\ref{eq:H-2mol})].

The zero-electron state is 
$|\psi^{(0)}\rangle =  |\, 0 \rangle$, and the corresponding energy 
  can be expressed in terms of model Hamiltonian as
$E^{(0)}=
\langle
\psi^{(0)}|H_{\mathrm{2\mbox{-}mol}}|\psi^{(0)}\rangle
=
(2 E_0  + \Delta E_0) - 3(\varepsilon_{\mathrm{g}} + \varepsilon_{\mathrm{u}})
+ 9(U_{\mathrm{g}} + U_{\mathrm{u}} + 4 U')/8 
+ 9 (V_{\mathrm{gg}} + V_{\mathrm{uu}} +2 V_{\mathrm{gu}})/4$.

By using Eq.\ (\ref{eq:correspondence}),
the one-electron states given in Eq.\ (\ref{eq:config-2mol-1el})
are re-expressed as
\begin{subequations}
\begin{eqnarray}
|\psi_{\mathrm{g,1}}^{(1)}\rangle &=&
\frac{1}{\sqrt{2}} (|\mathrm{g_\I}\rangle + |\mathrm{g_\II}\rangle),
\\ \nonumber \\
|\psi_{\mathrm{u,1}}^{(1)}\rangle &=&
\frac{1}{\sqrt{2}} (-|\mathrm{g_\I}\rangle + |\mathrm{g_\II}\rangle),
\\ \nonumber \\
|\psi_{\mathrm{g,2}}^{(1)}\rangle &=&
\frac{1}{\sqrt{2}} (-|\mathrm{u_\I}\rangle + |\mathrm{u_\II}\rangle),
\\ \nonumber \\
|\psi_{\mathrm{u,2}}^{(1)}\rangle &=&
\frac{1}{\sqrt{2}} (|\mathrm{u_\I}\rangle + |\mathrm{u_\II}\rangle),
\end{eqnarray}%
\label{eq:1el-2mol}%
\end{subequations}
where we have used the notation
  $|\mathrm{g_\I}\rangle=c_{\mathrm{g_1},\uparrow}^\dagger |0\rangle$
 and so on.

By focusing on these four one-electron states, 
the extended Hubbard model Hamiltonian (\ref{eq:H-2mol})
 can be expressed by the   $4\times 4$ matrix.
The molecular-based fragment MO  picture gives the
 following 
CI matrix expressed in the model space:
\renewcommand{\arraystretch}{1.5}
\begin{equation}
H_{\mathrm{2\mbox{-}mol}}
=
\left(
\begin{array}{cccc}
E_{\mathrm{g}}^{(1)} & - t_{\mathrm{gg}} & 
 \Delta \tilde\varepsilon_{\mathrm{gu}} & - t_{\mathrm{gu}}
\\ 
- t_{\mathrm{gg}} & E_{\mathrm{g}}^{(1)} &
 t_{\mathrm{gu}} & - \Delta \tilde\varepsilon_{\mathrm{gu}} 
\\ 
\Delta\tilde\varepsilon_\mathrm{gu} & t_\mathrm{gu} 
& E_{\mathrm{u}}^{(1)} & - t_{\mathrm{uu}} 
\\ 
- t_{\mathrm{gu}} & - \Delta \tilde\varepsilon_{\mathrm{gu}} 
& - t_{\mathrm{uu}} & E_{\mathrm{u}}^{(1)} 
\end{array}
\right)
\begin{array}{l}
|\mathrm{g_\I}\rangle \\ 
|\mathrm{g_\II}\rangle \\ 
|\mathrm{u_\I}\rangle \\ 
|\mathrm{u_\II}\rangle 
\end{array}
,
\end{equation}%
\renewcommand{\arraystretch}{1}%
where 
$E_\mathrm{g}^{(1)}
= 
(E^{(0)}
+   \varepsilon_{\mathrm{g}}
- 3 U_{\mathrm{g}}/4
- 3 U'/2
- 3 V_{\mathrm{gg}}/2
- 3 V_{\mathrm{gu}}/2
)$,
$E_\mathrm{u}^{(1)}
= 
(E^{(0)}
+   \varepsilon_{\mathrm{u}}
- 3 U_{\mathrm{u}}/4
- 3 U'/2
- 3 V_{\mathrm{uu}}/2
- 3 V_{\mathrm{gu}}/2
)$, and
$\Delta \tilde \varepsilon_{\mathrm{gu}}
=
(  \Delta \varepsilon_{\mathrm{gu}}
 + 3 X_{\mathrm{g}}/2
 + 3 X_{\mathrm{u}}/2 )
$.
The inter-molecular  transfer integrals $t_\mathrm{gg}$,
$t_\mathrm{uu}$, and $t_\mathrm{gu}$ 
 are shown in Fig.\ \ref{fig:model-gu}.
In terms of the basis given in Eq.\ (\ref{eq:1el-2mol}),
this CI matrix can be block diagonalized as
\begin{widetext}
\renewcommand{\arraystretch}{1.5}
\begin{equation}
H_{\mathrm{2\mbox{-}mol}}
=
\left(
\begin{array}{cccc}
E_{\mathrm{g}}^{(1)} - t_{\mathrm{gg}} & 
- \Delta \tilde\varepsilon_{\mathrm{gu}}- t_{\mathrm{gu}} & 
 0 & 0
\\ 
- \Delta \tilde\varepsilon_{\mathrm{gu}}- t_{\mathrm{gu}} & 
E_{\mathrm{u}}^{(1)} + t_\mathrm{uu} &
 0 & 0
\\ 
0 & 0 
& E_{\mathrm{g}}^{(1)} + t_{\mathrm{gg}} & 
- \Delta \tilde\varepsilon_{\mathrm{gu}} + t_{\mathrm{gu}} 
\\ 
0 & 0
& - \Delta \tilde\varepsilon_{\mathrm{gu}} + t_{\mathrm{gu}} 
 & E_{\mathrm{u}}^{(1)} -  t_\mathrm{uu} 
\end{array}
\right)
\begin{array}{l}
|\psi_{\mathrm{g,1}}^{(1)}\rangle \\
|\psi_{\mathrm{g,2}}^{(1)}\rangle \\
|\psi_{\mathrm{u,1}}^{(1)}\rangle \\
|\psi_{\mathrm{u,2}}^{(1)}\rangle
\end{array}
.
\label{eq:1el-2mo-2}
\end{equation}%
\renewcommand{\arraystretch}{1}%
\end{widetext}
Since the MOs are frozen to the ones obtained in the isolated molecule
calculation, the off-diagonal elements 
$\langle \psi_{\mathrm{g,1}}^{(1)} | H_{\mathrm{2\mbox{-}mol}} |
\psi_{\mathrm{g,2}}^{(1)} \rangle$ and
$\langle \psi_{\mathrm{u,1}}^{(1)} | H_{\mathrm{2\mbox{-}mol}} |
\psi_{\mathrm{u,2}}^{(1)} \rangle$ become 
non zero.

Similar implementations can be performed for the 2-electron case.
The expressions of each configuration 
  on the molecular-based fragment MO basis  and the corresponding 
  energies in terms of the extended Hubbard Hamiltonian 
 are shown in Table \ref{table:2mol2el}.
We also analyze the off-diagonal components, and
derive the CI matrix explicitly in terms of the parameters of 
 the extended Hubbard Hamiltonian.

\begin{table}[b]
\caption{
Estimated parameters for the neighboring 
 [Au(tmdt)$_2$]  molecules. All energies are in eV. 
}
\label{table:parameters-MO-Au-tmdt2}
\begin{ruledtabular}
\begin{tabular}{c|rrrrrr}
direction & $[100]$ & $[111]$  & $[101]$ & $[211]$ & $[001]$ & $[011]$ 
\\ \hline 
$t_{\mathrm{gg}}$
& $0.10$ 
& $0.10$
& $0.02$
& $0.01$
& $0.07$
& $0.01$
\\
$t_{\mathrm{uu}}$ 
& $0.12$ 
& $-0.19$
& $-0.05$
& $-0.02$
& $-0.11$
& $-0.02$
\\
$t_{\mathrm{gu}}$ 
& $0.00$ 
& $-0.14$
& $-0.03$
& $-0.01$
& $-0.09$
& $-0.02$
\\ \hline
$V_{\mathrm{gg}}$ 
& $1.75$ 
& $1.31$
& $1.38$
& $0.97$
& $1.37$
& $1.13$
\\
$V_{\mathrm{uu}}$ 
& $1.67$ 
& $1.39$
& $1.43$
& $1.00$
& $1.37$
& $1.14$
\\
$V_{\mathrm{gu}}$ 
& $1.70$ 
& $1.35$
& $1.41$
& $0.99$
& $1.38$
& $1.13$
\\ 
$I$ 
& $0.43$ 
& $-0.29$
& $-0.24$
& $-0.11$
& $-0.13$
& $-0.09$
\\ 
$X_\mathrm{g}$ 
& $0.09$ 
& $-0.56$
& $-0.52$
& $-0.28$
& $-0.45$
& $-0.31$
\\
$X_\mathrm{u}$ 
& $ 0.05$ 
& $-0.58$
& $-0.52$
& $-0.29$
& $-0.43$
& $-0.31$
\\ \hline
$\Delta\varepsilon_{\mathrm g}$  
& $ 0.26$ 
& $-0.11$
& $-0.06$
& $-0.07$
& $ 0.03$
& $-0.01$
\\
$\Delta\varepsilon_{\mathrm u}$  
& $ 0.20$ 
& $-0.06$
& $ 0.00$
& $-0.05$
& $ 0.07$
& $ 0.02$
\\
$\Delta \varepsilon_{\mathrm{gu}}$ 
& $-0.10$ 
& $ 0.21$
& $ 0.19$
& $ 0.04$
& $ 0.31$
& $ 0.12$
\end{tabular} 
\end{ruledtabular}
\end{table}

\subsection{Results}

Now we can evaluate the model parameters
 by relating the CI matrix obtained by the MR-CI calculations
  given in Eq.\ (\ref{eq:Habinitio}) and 
  the CI matrix expressed 
  in terms of the model Hamiltonian.
In the extended Hubbard Hamiltonian
  [Eq.\ (\ref{eq:H-2mol})], there are 
  17 parameters, 
  including the intra-molecular interactions.
In the two-electron case,
  22 diagonal terms and 30 independent off-diagonal terms 
can be determined
in the 
  \textit{ab initio} Hamiltonian (\ref{eq:Habinitio}).
Interestingly,
 the intra-molecular interactions, $U_\mathrm{g}$, $U_\mathrm{u}$, 
$U'$, and $J_\mathrm{H}$,
are also obtained 
from the calculations for the  two-molecule system.
These values are consistent with the results in Table 
  \ref{eq:parameter-1mol}.

For [Au(tmdt)$_2$],
the inter-molecular interactions 
are listed in Table \ref{table:parameters-MO-Au-tmdt2}.
From the data of transfer integrals,
we see that the system exhibits three-dimensional character, 
 in agreement with  DFT-based calculations. \cite{Ishibashi2008}
The estimated parameters for $\Delta \varepsilon_\mathrm{g}$,
 $\Delta \varepsilon_\mathrm{u}$, and 
 $\Delta \varepsilon_\mathrm{gu}$ 
 are also listed in Table
\ref{table:parameters-MO-Au-tmdt2} 
and are comparable to the energy
  difference 
 $(\varepsilon^0_\mathrm{g}-\varepsilon^0_\mathrm{u})\approx 0.26$ eV.
For (TTM-TTP)I$_3$, 
the inter-molecular interactions 
exhibit strong anisotropy,
 i.e.,
the inter-molecular interactions 
  for the [001] molecule pair become largest compared with 
  those for other pairs.\cite{SI}
These features can be explained by noting that
   [001] is the stacking direction  of TTM-TTP molecules 
  and the two-molecule distance becomes much shorter for this 
  face-to-face  molecule pair. 
The parameters $\Delta \varepsilon_\mathrm{g}$ and 
$\Delta \varepsilon_\mathrm{u}$
for the ionic TTM-TTP$^+$ molecules are large in contrast to 
those for the neutral [Au(tmdt)$_2$] molecules.
Since these values are almost identical $\Delta \varepsilon_\mathrm{g} 
   \simeq \Delta \varepsilon_\mathrm{u}$, we find that 
 the g and u MOs are still 
 quasi-degenerate in the crystal.

It is known that 
the orbital exchange interaction term $I$ 
accounts for dispersion interactions
(i.e., van der Waals interactions) between pairs of molecules.
\cite{Vincent_review,London:1930uz,Perraud:2011cz}
These correspond to instantaneous dipole-dipole interactions resulting
 from the local charge excitations $\mathrm{g}_\I \to \mathrm{u}_\I$ and 
 $\mathrm{g}_\II \to \mathrm{u}_\II$
[see Eq.\ (\ref{eq:I-2mol})], with the excitation energy $\Delta E$.
The dispersion interaction can be evaluated as
  $\sim -I^2/\Delta E$, and 
the typical amplitudes of the dispersion interaction are 
in the range $-0.1$ to  $-0.01$ eV.
In the present systems, the dispersion interactions
for nearest-neighboring molecules are 
$-0.24$ eV for [Au(tmdt)$_2$]$_2$ and
$-0.08$ eV for [TTM-TTP$^-$]$_2$.
For the typical one-band system of the TTF molecules,
 we also evaluated the parameter $I=0.19$ eV and 
 found that the dispersion interaction is very weak ($\sim -0.006$ eV).
Such an observation can be ascribed to the fact that 
the present systems are very polarizable due to extended MOs.

In the way shown above,
 we  determine all the possible model parameters uniquely for
  multi-orbital systems by 
 taking the advantage of the
  wavefunction-based \textit{ab initio} calculations.
However, the resulting parameters shown here are bare values and 
the  screening effects are not taken into account.
A framework to project the effective model from large CI expansions 
 has been developed
where  
the electronic excitations involving
  valence and virtual MOs, i.e.,
  so-called dynamic correlations, are taken into account.
\cite{Bloch1958,Calzado2002a,Calzado2002b}
With this scheme, 
   one can access to accurate  magnetic interactions, and simultaneously,
   transfer integral $t$, Coulomb repulsion $U$, and the 
  direct exchange coupling, with the inclusion of 
  these screening effects.
For \textit{atomic} orbitals in metal complexes, 
 the screening effects to Coulomb repulsions are pronounced 
   due to the localized orbital nature and also due to the 
  presence of ligands, 
typically 
  by a factor of 4 to 5.
 \cite{Calzado2002b}
It has been shown that this scheme is relevant to 
the analysis of pure organic materials, 
\cite{Calzado:2010cw,Verot2011}
and the screening effects would be less important for 
 larger molecules.\cite{Cano-Cortes2007}
From the recent approaches based on DFT,
  where the \textit{inter-molecular} screening effects 
 in addition to the above \textit{intra-molecular} screening effects
  are taken into account,
 the bare magnitudes for the Coulomb repulsions 
are reduced 
to at most a quarter.
\cite{Cano-Cortes2007,Nakamura2009}

\newpage \phantom{a}

\section{Model for crystal system and band structure}

In this section, we consider the crystal system 
 and evaluate the band structure.
The symmetry-breaking term $\Delta \varepsilon_{\mathrm{gu}}$
  [Eq.\ (\ref{eq:H-Delta})]
  disappears in the periodic crystal system. 
The full Hamiltonian is given by
  $H_{\mathrm{cryst}}=H^{\mathrm{kin}}_{\mathrm{cryst}}
                       +H^{\mathrm{int}}_{\mathrm{cryst}}$ with
\begin{widetext}
\begin{subequations}
\begin{eqnarray}
H^{\mathrm{kin}}_{\mathrm{cryst}}
&=& 
\sum_j 
(
  \varepsilon_{\mathrm g} n_{\mathrm g,j} 
+ \varepsilon_{\mathrm u} n_{\mathrm u,j} 
)
-  \sum_{\langle i,j \rangle}\sum_\sigma 
 t_{\mathrm{gg}}^{[n_a,n_b,n_c]}
(c_{\mathrm{g},i,\sigma}^\dagger c_{\mathrm{g},j,\sigma}^{} +\mathrm{h.c.})
-  \sum_{\langle i,j \rangle} \sum_\sigma 
  t_{\mathrm{uu}} ^{[n_a,n_b,n_c]}
(c_{\mathrm{u},i,\sigma}^\dagger c_{\mathrm{u},j,\sigma}^{} +\mathrm{h.c.})
\nonumber \\ && {}
-  \sum_{\langle i,j \rangle}  \sum_\sigma 
t_{\mathrm{gu}}^{[n_a,n_b,n_c]}
(c_{\mathrm{g},i,\sigma}^\dagger c_{\mathrm{u},j,\sigma}^{} 
-c_{\mathrm{u},i,\sigma}^\dagger c_{\mathrm{g},j,\sigma}^{} +\mathrm{h.c.})
,
\label{eq:H-crystal-kin}
\\ 
H^{\mathrm{int}}_{\mathrm{cryst}}
&=& 
\sum_j 
\biggl\{
   U_{\mathrm g} n_{\mathrm g,j,\uparrow} n_{\mathrm g,j,\downarrow}
 + U_{\mathrm u} n_{\mathrm u,j,\uparrow} n_{\mathrm u,j,\downarrow}
 + U' n_{\mathrm g,j} n_{\mathrm u,j}
-J_{\mathrm H} 
\left[
\bm{S}_{\mathrm g,j} \cdot \bm{S}_{\mathrm u,j}
-\frac{1}{2}
 (c_{\mathrm{g},j,\uparrow}^\dagger c_{\mathrm{u},j,\uparrow}^{} 
  c_{\mathrm{g},j,\downarrow}^\dagger c_{\mathrm{u},j,\downarrow}^{}
   + \mathrm{h.c.})
\right]
\biggr\}
\nonumber \\ && {}
+  \sum_{\langle i,j \rangle}
       V_{\mathrm{gg}}^{[n_a,n_b,n_c]}  n_{\mathrm{g},i} n_{\mathrm{g},j}
 + \sum_{\langle i,j \rangle}
       V_{\mathrm{uu}}^{[n_a,n_b,n_c]}  n_{\mathrm{u},i} n_{\mathrm{u},j}
 + \sum_{\langle i,j \rangle}
      V_{\mathrm{gu}}^{[n_a,n_b,n_c]}
       \left( n_{\mathrm{g},i} n_{\mathrm{u},j} 
            + n_{\mathrm{u},i} n_{\mathrm{g},j} \right)
\nonumber\\
&&
+  
 \sum_{\langle i,j \rangle} \sum_{\sigma,\sigma'}  
I^{[n_a,n_b,n_c]}
 \left(
   c_{\mathrm{g},i,\sigma}^\dagger   c_{\mathrm{u},i,\sigma} 
   c_{\mathrm{u},j,\sigma'}^\dagger  c_{\mathrm{g},j,\sigma'} 
 + c_{\mathrm{g},i,\sigma}^\dagger   c_{\mathrm{u},i,\sigma} 
   c_{\mathrm{g},j,\sigma'}^\dagger  c_{\mathrm{u},j,\sigma'} 
 + \mathrm{h.c.}
\right)
\nonumber\\
&& {}
+     
\sum_{\langle i,j \rangle} \sum_\sigma
X_\mathrm{g}^{[n_a,n_b,n_c]}
\left[
 n_{\mathrm{g},i} \,
( c_{\mathrm{g},j,\sigma}^\dagger c_{\mathrm{u},j,\sigma} + \mathrm{h.c.} )
-
( c_{\mathrm{g},i,\sigma}^\dagger c_{\mathrm{u},i,\sigma} + \mathrm{h.c.} )
\, n_{\mathrm{g},j}
\right]
\nonumber \\ && {}
+
\sum_{\langle i,j \rangle} \sum_\sigma
X_\mathrm{u}^{[n_a,n_b,n_c]}
\left[
 n_{\mathrm{u},i} \, 
( c_{\mathrm{g},j,\sigma}^\dagger c_{\mathrm{u},j,\sigma} + \mathrm{h.c.} )
- 
( c_{\mathrm{g},i,\sigma}^\dagger c_{\mathrm{u},i,\sigma} + \mathrm{h.c.} )
\, n_{\mathrm{u},j} 
\right],
\label{eq:H-crystal-int}%
\end{eqnarray}%
\label{eq:H-crystal}%
\end{subequations}
\end{widetext}
where $\langle i, j \rangle$ denotes the combination
  of neighboring molecule pair and is assumed that   
 $i$ represents the molecule in a reference  position 
  while $j$ the translated molecule by 
  the vector [$n_a,n_b,n_c$] shown in Fig.\ \ref{fig:crystal}. 

\begin{figure}[t]
\includegraphics[width=8.cm,bb=0 0 312 228]{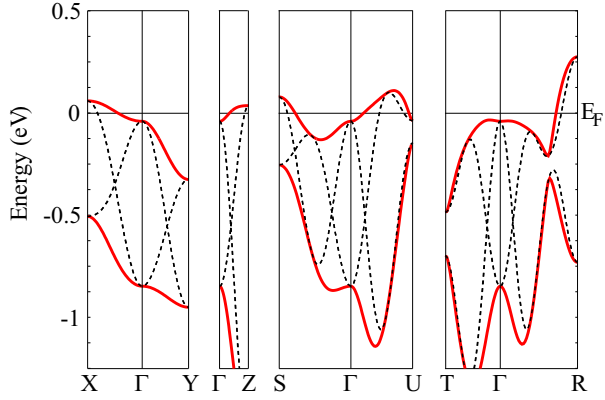}
\caption{
Band structure of 
[Au(tmdt)$_2$]
obtained
  from a parametrization based on our \textit{ab initio} calculations,
where $\Gamma=(0,0,0)$, $\mathrm X=(\pi,0,0)$, $\mathrm Y=(0,\pi,0)$, 
$\mathrm Z=(0,0,\pi)$, $\mathrm S=(\pi,\pi,0)$, $\mathrm U=(\pi,0,\pi)$, 
$\mathrm T=(0,\pi,\pi)$, and $\mathrm R=(\pi,\pi,\pi)$.
The dotted curves represent the energy dispersion setting
  $t_{\mathrm{gu}}=0$.
}
\label{fig:band}
\end{figure}

Next we examine the band structure 
for crystals by neglecting the correlation effects.
In order to determine the energy levels,
the crystal field effect arising from the 
$\Delta \varepsilon_\mathrm{g}$ and $\Delta \varepsilon_\mathrm{u}$ terms
must be taken into account.
We note that, for (TTM-TTP)I$_3$,
the potential due to the counterion I$_3^-$
should also be included
in order to examine this crystal field effect quantitatively.
In addition,
the Hartree corrections from Eq.\ (\ref{eq:H-crystal-int})
also contribute to the energy difference of MO level energies; however,
these contributions would be small since 
the density operators in Hamiltonian 
  are represented in the normal-ordered form [see Eq.\ (\ref{eq:densityop})].
In the present analysis for [Au(tmdt)$_2$], 
we assign the MO level energies as
$\varepsilon_\mathrm{g} \approx
\varepsilon_\mathrm{g}^0 + 2 \Delta \varepsilon_\mathrm{g}^\mathrm{[100]}$ and
$\varepsilon_\mathrm{u} \approx
\varepsilon_\mathrm{u}^0 + 2 \Delta \varepsilon_\mathrm{u}^\mathrm{[100]}$,
where the factor 2 
reflects
the coordination number, and 
we simply focus on 
  the kinetic term [Eq.\ (\ref{eq:H-crystal-kin})].
The resulting band structure 
for the [Au(tmdt)$_2$] crystal
is shown in Fig.\ \ref{fig:band}.
The bandwidth obtained from the present 
  analysis is overestimated in comparison with the
result of DFT-based calculation for the periodic system, \cite{Ishibashi2005}
however, 
qualitative behavior of the band structure 
 is well reproduced.
By taking advantage of the present scheme,
we can elucidate 
 the nature of the
 band structure, 
by setting the mixing term to zero, i.e., $t_{\mathrm{gu}}= 0$. 
Such a fictitious band structure is also shown in 
 Fig.\ \ref{fig:band}.
We observe that 
the two bands overlap and mix in together
by large mixing amplitude,  supporting a multi-band system.

The band structure for (TTM-TTP)I$_3$ is also analyzed.\cite{SI}
From an extended H\"uckel approach, 
 \cite{Mori1984_Huckel}
the overlap integrals between the neighboring molecules
  along the stacking direction were given by
  $S_\mathrm{gg}=-0.19\times 10^{-3}$, 
  $S_\mathrm{uu}=-26.22\times 10^{-3}$, and
  $S_\mathrm{gu}=-12.92\times 10^{-3}$.
The small transfer integral $t_{\mathrm{gg}}$ and also 
  the small overlap integral $S_{\mathrm{gg}}$ 
result from   the tilted
  alignment of TTM-TTP molecules
  along the stacking direction in the crystal. 
We observe that the band built on the u MO is much wider 
  than the one built on the g MO. 
The bandwidth for the u MO is $\sim 1.2$ eV while that for the g MO 
is $\sim 0.3$ eV.
Finally, 
since the inter-orbital transfer integral
 $t_\mathrm{gu}^{[001]}$ ($=-0.13$ eV) 
  is 
 relatively large compared with $t_\mathrm{gg}$, 
   the g MO band is strongly modified,
while  the u band is not much affected 
near the Fermi energy.

\begin{figure*}[t]
\includegraphics[width=10cm,bb=0 0 552 256]{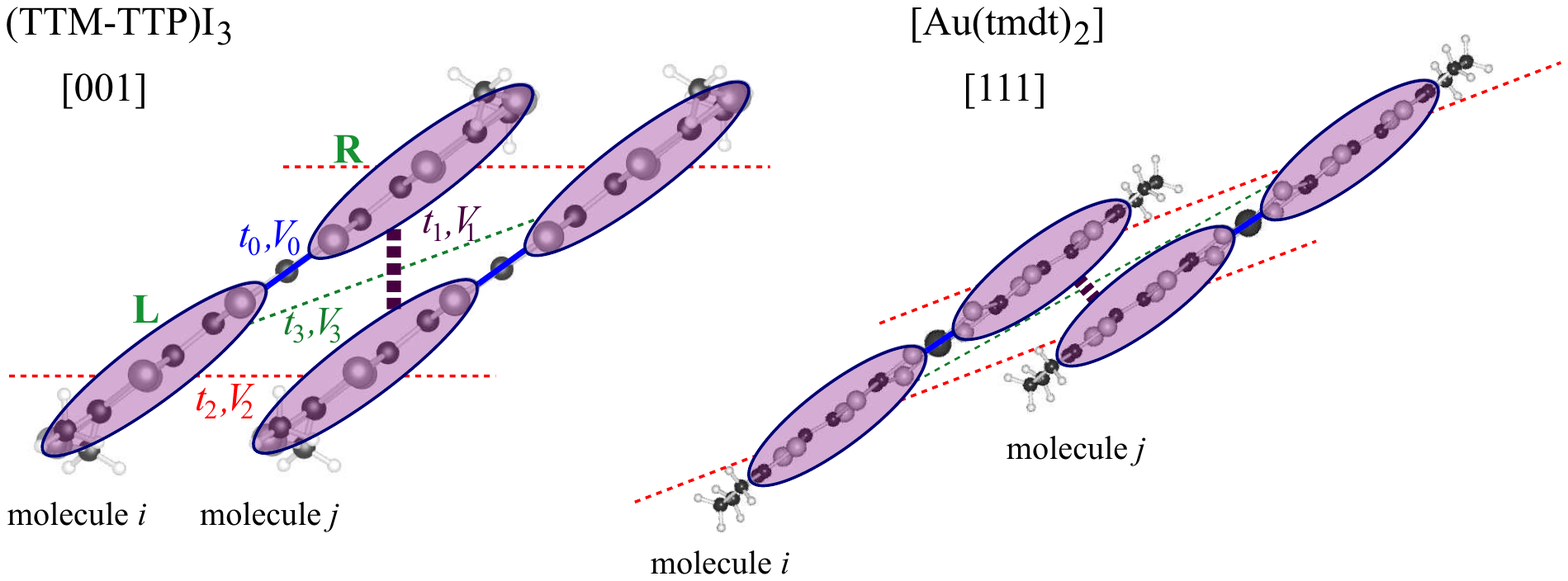}
\caption{
The full-fragment-MO picture and 
the definition of parameters for (TTM-TTP)I$_3$ (left) 
and [Au(tmdt)$_2$] (right).
}
\label{fig:fragment}
\end{figure*}

\section{Full fragment decomposition}

In this section, we derive the effective model 
  based on full fragment MOs, 
  which is the most fundamental model to 
  analyze the ``intra-molecular'' degree of freedom. 
The relevant fragment MOs are simply  the left and right 
 part of MOs,  $\varphi_\mathrm{L}$ 
 and $\varphi_\mathrm{R}$,
  where the center fragment is omitted 
  due to its small weight. 
\cite{Marie2010}
We denote the corresponding annihilation operators by 
  $c_{\mathrm{L},j,\sigma}$  and   $c_{\mathrm{R},j,\sigma}$.
The operator correspondence between the original MOs
  and the fragment MOs basis is given by 
\begin{subequations}
\begin{eqnarray}
c_{\mathrm g,j,\sigma} 
&=& 
\frac{1}{\sqrt{2}} 
\left(
-  c_{\mathrm L,j,\sigma} + c_{\mathrm R,j,\sigma}
\right)
,
\\ \nonumber \\
c_{\mathrm u,j,\sigma} 
&=& 
\frac{1}{\sqrt{2}} 
\left(
  c_{\mathrm L,j,\sigma} + c_{\mathrm R,j,\sigma}
\right)
.
\end{eqnarray}
\end{subequations}
Let us mention that  $\varphi_{L}$ and $\varphi_{R}$ are not orthogonal,  but
 the overlap integral is very small.
 \cite{Marie2010}

On the fragment MO basis,
 the model Hamiltonian for the crystal [Eq.\ (\ref{eq:H-crystal})]
can be re-expressed as
  $H_{\mathrm{cryst}}=H^{\mathrm{intramol}}_{\mathrm{cryst}}
                       +H^{\mathrm{intermol}}_{\mathrm{cryst}}$ with
\begin{eqnarray}
&&
\hspace*{-.5cm}
H_{\mathrm{crystal}}^{\mathrm{intramol}}
\nonumber \\
&=&  
\varepsilon_0 \sum_{j}
( 
  n_{\mathrm L,j} + n_{\mathrm R,j} 
)
\nonumber \\ && {} \hspace*{-.5cm}
- t_0 
\sum_{j} \sum_\sigma
( 
    c_{\mathrm L,j,\sigma}^\dagger c_{\mathrm R,j,\sigma}
  + c_{\mathrm R,j,\sigma}^\dagger c_{\mathrm L,j,\sigma}  
)  
\nonumber \\ && {}\hspace*{-.5cm}
+   U \sum_{j}
   \left(
     n_{\mathrm{L},j,\uparrow}
     n_{\mathrm{L},j,\downarrow}
   + n_{\mathrm{R},j,\uparrow}
     n_{\mathrm{R},j,\downarrow}
   \right)
\nonumber \\ && {}\hspace*{-.5cm}
+ V_0 \sum_{j} 
     n_{\mathrm{L},j}
     n_{\mathrm{R},j}
\nonumber \\ && {}\hspace*{-.5cm}
 - J \sum_{j}  \!\!
   \left[
      \bm{S}_{\mathrm{L},j} \cdot \bm{S}_{\mathrm{R},j}
  - \frac{1}{2}
   (c_{\mathrm{L},j,\uparrow}^\dagger \,
    c_{\mathrm{L},j,\downarrow}^\dagger \, 
    c_{\mathrm{R},j,\downarrow} \, 
    c_{\mathrm{R},j,\uparrow} + \mathrm{h.c.})
   \right]
\nonumber \\ && {}\hspace*{-.5cm}
 + X \sum_{j,\sigma} 
   (n_{\mathrm{L},j,\sigma} + n_{\mathrm{R},j,\sigma} )
   (c_{\mathrm{L},j,\tilde\sigma}^\dagger 
    c_{\mathrm{R},j,\tilde\sigma} + \mathrm{h.c.} )
,
\\ \nonumber \\
&& 
\hspace*{-.5cm}
H_{\mathrm{crystal}}^{\mathrm{intermol}}
\nonumber \\
&=&  
- \sum_{\langle i,j \rangle}
 \sum_\sigma
 t_1^{[n_a,n_b,n_c]}
( 
    c_{\mathrm R,i,\sigma}^\dagger c_{\mathrm L,j,\sigma} + \mathrm{h.c.}
)  
\nonumber \\ && {}
-\sum_{\langle i,j \rangle}
 \sum_\sigma
  t_2^{[n_a,n_b,n_c]}
( 
    c_{\mathrm R,i,\sigma}^\dagger c_{\mathrm R,j,\sigma}
  + c_{\mathrm L,i,\sigma}^\dagger c_{\mathrm L,j,\sigma}
  + \mathrm{h.c.}
)  
\nonumber \\ && {} 
- \sum_{\langle i,j \rangle}
 \sum_\sigma
  t_3^{[n_a,n_b,n_c]}
( 
    c_{\mathrm L,i,\sigma}^\dagger c_{\mathrm R,j,\sigma} + \mathrm{h.c.}
)
\nonumber \\ && {}
+ \sum_{\langle i,j\rangle}
  V_1^{[n_a,n_b,n_c]}
      n_{\mathrm{R},i}
      n_{\mathrm{L},j}
\nonumber \\ && {}
+ \sum_{\langle i,j\rangle}
  V_2^{[n_a,n_b,n_c]} 
      \left(
      n_{\mathrm{L},i}
      n_{\mathrm{L},j}
    + n_{\mathrm{R},i}
      n_{\mathrm{R},j}
      \right)
\nonumber \\ && {}
+ \sum_{\langle i,j\rangle}
  V_3^{[n_a,n_b,n_c]} 
      n_{\mathrm{L},i}
      n_{\mathrm{R},j}
,
\end{eqnarray}
where $\tilde\sigma=\uparrow(\downarrow)$ for $\sigma=\downarrow(\uparrow)$.
The parameters $\varepsilon_0$ and $t_0$ represent
  the energy level of the fragment MO and the inter-fragment transfer
  integral within the molecule.
The parameter $U$ represents the magnitude of the Coulomb repulsion 
 between electrons  within the fragment MOs,  and 
$V_0$, $J$, and $X$ denote the inter-fragment Coulomb repulsion,
  exchange interaction, and bond-density interactions 
  within the molecule, respectively.
The inter-molecular transfer integrals
 $t_i$ $(i=1,2,3)$ and the inter-molecular Coulomb repulsions
 $V_i$ $(i=1,2,3)$ are depicted in Fig.\ \ref{fig:fragment}.  
The density operators are given in the normal-ordered form
$n_{\mathrm{L/R},j,\sigma}
=
(c_{\mathrm{L/R},j,\sigma}^\dagger c_{\mathrm{L/R},j,\sigma}^{}-\frac{3}{4})$
and the basis-set change leads to
\begin{subequations}
\begin{eqnarray}
\varepsilon_0 &=& 
\frac{1}{2} (\varepsilon_{\mathrm{g}} + \varepsilon_{\mathrm{u}}) , 
\\ \nonumber \\
t_0 &=& 
\frac{1}{2} (\varepsilon_{\mathrm{g}}
  - \varepsilon_{\mathrm{u}}),
\\ \nonumber \\
U&=& \frac{1}{4}  (U_{\mathrm{g}} + U_{\mathrm{u}}) 
+ \frac{1}{2}U' + \frac{5}{8} J_\mathrm{H},
\\ \nonumber \\
V_0 &=& \frac{1}{8} (U_{\mathrm{g}} + U_{\mathrm{u}} )
+ \frac{3}{4}U' - \frac{5}{16} J_\mathrm{H},
\\ \nonumber \\
J &=&
\frac{1}{2} (U_{\mathrm{g}} + U_{\mathrm{u}} )
 -U'  - \frac{1}{4} J_\mathrm{H},
\\ \nonumber \\
X &=& \frac{1}{4} (- U_{\mathrm{g}} + U_{\mathrm{u}}),
\end{eqnarray}
\end{subequations}
for the intra-molecular parameters.
This kind of transformation was applied 
for simpler two-orbital systems.
\cite{BonacicKoutecky:1987uu}
Similarly, the inter-molecular interactions can be expressed as
\begin{subequations}
\begin{eqnarray}
t_1 &=&
\frac{1}{2} (- t_{\mathrm{gg}} + t_{\mathrm{uu}} + 2 t_{\mathrm{gu}}),
\\ \nonumber \\
t_2 &=&
\frac{1}{2} (t_{\mathrm{gg}} + t_{\mathrm{uu}} ),
\\ \nonumber \\
t_3 &=&
\frac{1}{2} (- t_{\mathrm{gg}} + t_{\mathrm{uu}} - 2 t_{\mathrm{gu}}),
\\ \nonumber \\
V_1 
&=&
\frac{1}{4} (V_{\mathrm{gg}} + V_{\mathrm{uu}} )
+ \frac{1}{2}V_{\mathrm{gu}} 
- I
- X_\mathrm{g}
- X_\mathrm{u},
\\ \nonumber \\
V_2 
&=&
\frac{1}{4} (V_{\mathrm{gg}} + V_{\mathrm{uu}} )
+ \frac{1}{2}V_{\mathrm{gu}}
+ I,
\\ \nonumber \\
V_3 
&=&
\frac{1}{4} (V_{\mathrm{gg}} + V_{\mathrm{uu}} )
+ \frac{1}{2}V_{\mathrm{gu}} 
- I
+ X_\mathrm{g}
+ X_\mathrm{u},
\qquad
\end{eqnarray}
\end{subequations}
where the index $[n_a,n_b,n_c]$ is suppressed.
The evaluated parameters are summarized in Table
\ref{table:parameter-crystal-fragment}.
It is worthwhile to note that  the presence of
  the nontrivial orbital exchange interaction $I$ and 
the  bond-charge interaction $X_\mathrm{g}$ and $X_\mathrm{u}$ plays
 a crucial role to differentiate 
 the Coulomb interactions, $V_1$, $V_2$, and $V_3$
   in the fragment-MO picture.

\begin{table}[b]
\caption{
Estimated parameters for the inter-molecular interactions 
  on the full-fragment MO basis for (TTM-TTP)I$_3$.
The intra-molecular interactions are given by
$\varepsilon_0=-12.07$ eV,
 $t_0=-0.17$ eV,
  $U=5.30$ eV, $V_0=2.07$ eV, $J=0.18$ eV, and $X=0.05$ eV.
All energies are in eV.
}
\label{table:parameter-crystal-fragment}
\begin{ruledtabular}
\begin{tabular}{c|rrrrrrr}
 & $[001]$ & $[002]$ & $[012]$  & $[013]$ & $[100]$ & $[10\bar{1}]$ &
 $[10\bar{2}]$ 
\\\hline 
  $t_1$ 
&  $-0.26$ 
&  $-0.01$
&  $-0.05$
&  $0.02$
&  $0.00$
&  $0.00$
&  $0.00$
\\
  $t_2$ 
&  $-0.17$ 
&  $0.00$
&  $0.01$
&  $0.00$
&  $0.00$
&  $0.00$
&  $0.00$
\\
  $t_3$ 
&  $0.01$ 
&  $0.00$
&  $0.00$
&  $0.00$
&  $0.00$
&  $-0.01$
&  $0.00$
\\ \hline
  $V_1$
&  $2.67$ 
&  $1.88$
&  $1.87$
&  $1.36$
&  $0.86$
&  $0.71$
&  $0.58$
\\
  $V_2$
&  $2.47$ 
&  $1.27$
&  $0.99$
&  $0.77$
&  $1.28$
&  $1.08$
&  $0.83$
\\
  $V_3$
&  $1.23$ 
&  $0.79$
&  $0.67$
&  $0.53$
&  $1.46$
&  $1.63$
&  $1.30$
\end{tabular} 
\end{ruledtabular}
\end{table}

\begin{table}[b]
\caption{
Estimated parameters for the inter-molecular interactions 
  on the full-fragment MO basis for [Au(tmdt)$_2$].
The intra-molecular interactions are given by
$\varepsilon_0=-5.70$ eV,
 $t_0=0.19$ eV,
  $U=5.26$ eV, $V_0=1.58$ eV, $J=0.08$ eV, and $X=-0.01$ eV.
All energies are in eV.
}
\begin{ruledtabular}
\begin{tabular}{c|rrrrrr}
 & $[100]$ & $[111]$  & $[101]$ & $[211]$ & $[001]$ & $[011]$ 
\\\hline 
  $t_1$ 
&   $0.01$ 
&  $-0.29$
&  $-0.07$
&  $-0.03$
&  $-0.18$
&  $-0.04$
\\
  $t_2$ 
&   $0.11$ 
&  $-0.04$
&  $-0.02$
&  $ 0.00$
&  $-0.02$
&  $ 0.00$
\\
  $t_3$ 
&  $ 0.01$ 
&  $ 0.00$
&  $ 0.00$
&  $ 0.00$
&  $ 0.00$
&  $ 0.00$
\\ \hline
  $V_1$
&  $1.13$ 
&  $2.79$
&  $2.69$
&  $1.67$
&  $2.39$
&  $1.85$
\\
  $V_2$
&  $2.14$ 
&  $1.06$
&  $1.16$
&  $0.87$
&  $1.24$
&  $1.04$
\\
  $V_3$
&  $1.41$ 
&  $0.50$
&  $0.62$
&  $0.54$
&  $0.61$
&  $0.60$
\end{tabular} 
\end{ruledtabular}
\end{table}
   
\begin{figure}[t]
\includegraphics[width=6cm,bb= 45 243 443 619]{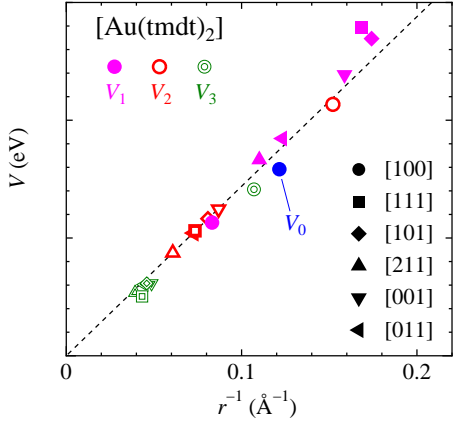}
\caption{
The Coulomb repulsions
as a function of inverse inter-fragment distance $r^{-1}$
for [Au(tmdt)$_2$] .
The results for (TTM-TTP)I$_3$ are shown in the supproting information.\cite{SI}
The filled, open and double symbols represent $V_1$, $V_2$, and $V_3$,
respectively.
The repulsion between the L and R fragments
 within the molecule is characterized by $V_0$.
The Coulomb repulsion within the fragment is 
$U\simeq 5.26$ eV.
The dotted line represents the bare Coulomb repulsion 
$a/r$ with  $a\approx 14.4$ eV \AA.
}
\label{fig:rdep-frag}
\end{figure}

The magnitudes of the Coulomb repulsion within the fragment, 
 $U\simeq 5.30$ eV for TTM-TTP$^+$ 
and  $U\simeq 5.26$ eV for [Au(tmdt)$_2$],
 are comparable with that for the 
  TTF molecule $U_\mathrm{TTF}\simeq 6.2 $ eV.
For the inter-fragment transfer integrals, 
not only the nearest-neighbor Coulomb
 interactions but the long-range interactions have large amplitudes.
In order to verify the fragment decomposition, we 
  examine the distance dependence of the inter-fragment interactions.
 We define the inter-fragment distance $r$ 
  by the averaged inverse distance between the $N_\mathrm{S}=6$
sulfur atoms in 
  each fragment, by
\begin{eqnarray}
\frac{1}{r} = \frac{1}{N_\mathrm{S}^2}\sum_{ij}
  \frac{1}{r_{ij}},
\end{eqnarray}
where $r_{ij}$ is the distance between the sulfur atoms.
The $r$-dependences of the inter-fragment interaction $V$ 
for [Au(tmdt)$_2$] and  (TTM-TTP)I$_3$ 
are
  shown in Fig.\ \ref{fig:rdep-frag}
and in the 
supporting information.\cite{SI}
The dotted curve denotes
the bare Coulomb  interactions  $V(r)=1/(4\pi \varepsilon_0 r)$,
 where  $\varepsilon_0$ is the permittivity of vacuum.
Thus, we conclude that $V$ follows 
  well the Coulomb law, and 
  our fragment decomposition 
is verified from this evaluation 
  of inter-fragment Coulomb repulsion.
Incidentally, the fact that the intra-molecular Coulomb repulsion $V_0$
  also follows well the Coulomb repulsion 
supports that the center fragment can be neglected.

From the data of inter-fragment transfer integrals, we find 
  that the stacking TTM-TTP molecules ([001] direction) 
 can be described 
  as a two-leg ladder system, where 
  the transfer integral along leg direction is $t_2$, 
  while those along the rung direction are $t_0$ and $t_1$.
It is worth noting that 
the inter-molecular transfer integral $t_1$ exceeds
  the intra-molecular transfer integral $t_0$.
This implies that, as a possible origin to the insulating behavior 
  at high temperature, electrons are localized on each $t_1$ bond,
a reminiscence of
 the ``dimer-Mott'' state in 
  the quarter-filled 1D 
systems. \cite{Seo1997,Tsuchiizu2001,Tsuchiizu2001b}

\begin{figure}[b]
\includegraphics[width=8cm,bb=0 0 396 320]{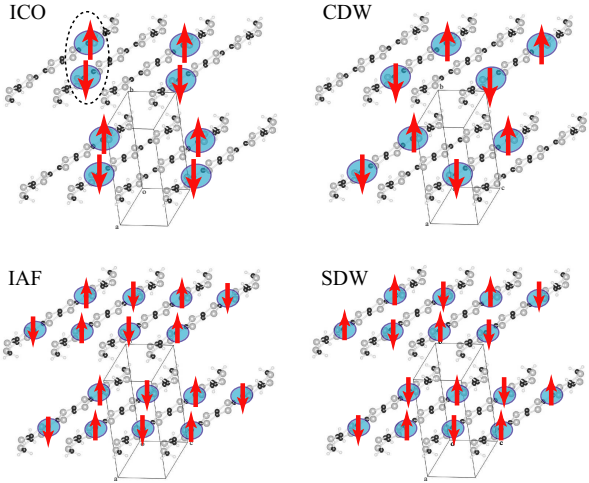}
\caption{
Strong-coupling picture for several possible ordered states in (TTM-TTP)I$_3$
shown on the fragment-MO basis.
Each circle (arrow) in the ICO and charge-density-wave (CDW)
 states represents one  hole
 (spin), 
while that in the IAF and spin-density-wave (SDW)
  states represents  half of hole (spin).
In the ICO state, 
two holes surrounded by the dotted circle 
 represent 
a spin-singlet pair.
In the CDW state, two holes form a spin-singlet pair within a molecule.
}
\label{fig:fig8}
\end{figure}

Here we  discuss the low-temperature symmetry-broken states  
  in (TTM-TTP)I$_3$ 
  on the basis of the present model parameters.
In  (TTM-TTP)I$_3$,
  the intra-molecular charge ordering has been proposed
from the Raman and x-ray experiments.
\cite{Yakushi2003,Swietlik2004,Swietlik2005,Nogami2003} 
However, 
  the charge pattern and the origin of the spin-singlet behavior 
 have not been clarified yet.
From a simple strong-coupling
 analysis,
we can examine the energies of possible ordered states 
  as shown in Fig.\ \ref{fig:fig8}.
We observe that the lowest-energy
state is the ICO state 
($E_\mathrm{ICO}=-1.47$ eV per molecule)
in which the charge is disproportionated
within each constitutive molecule.
This ICO
pattern is compatible with the $q_1=(0,0,1/2)$ superstructure
observed in the x-ray
  measurements. \cite{Maesato1999,Fujimura1999,Nogami2003}
From  the evaluated model parameters, 
  we expect that the super-exchange interaction 
  along the $t_1$ bond becomes largest and it would play a 
  crucial role to induce the spin-singlet state.
 On the basis
of this finding, we infer that the non-magnetic insulating
behavior  observed at low temperatures  in  (TTM-TTP)I$_3$
  \cite{Maesato1999,Fujimura1999,Onuki2001_Synth,Onuki2001_JPCS}
can be attributed to
the spin-singlet formation on the $t_1^{\mathrm{[001]}}$
bond 
(shown by the dotted circle in Fig.\ \ref{fig:fig8})
with twofold periodicity along the stacking direction. 
Incidentally, 
we observe that the IAF state is almost degenerate with the ICO state, 
with 
the energy differences per molecule 
$E_{\mathrm{IAF}}-E_{\mathrm{ICO}}\approx 0.15$ eV, 
$E_{\mathrm{SDW}}-E_{\mathrm{ICO}}\approx 0.30$ eV, and 
$E_{\mathrm{CDW}}-E_{\mathrm{ICO}}\approx 1.19$ eV.
Detailed analysis of possible symmetry-broken states  
  described by the present Hamiltonian 
  has been reported by using the mean-field approximation. \cite{Omori_MF}

From the optical measurement analysis,
a nontrivial optical absorption band has been observed 
at $\approx$  5000 cm$^{-1}$,
corresponding to a charge transfer band.\cite{Swietlik2005}
It has been suggested that 
this band can only be observed 
under some electric fields polarized perpendicular to the
  stacking direction $E\perp c$.  Furthermore 
its intensity 
is strongly enhanced
at low temperatures.
A possible scenario to 
explain this behavior 
is the following.
In the ICO state, 
a characteristic charge excitation 
  perpendicular to the stacking direction 
  can be described with two holes localized on two adjacent 
  fragments (see 
  dotted circle in Fig.\ \ref{fig:fig8}).
If one considers a two-site two-electron system as a simplest model for
this unit,  the ground state is singlet with 
energy  
  $E_0=(U+V_1^{[001]})/2-[(U-V_1^{[001]})^2/4+4(t_1^{[001]})^2]^{1/2}$.
While the first-excited state with energy
$E_1=V_1$ represents an optically-forbidden spin transition,
 the unique optically-allowed transition 
involves the 
second-excited state 
with energy $E_2=U$, representing a charge excitation.
The corresponding excitation 
 energy can be roughly estimated as 6800 cm$^{-1}$ by using 
  reduced Coulomb values
(by a factor of 5).
\cite{Cano-Cortes2007,Nakamura2009}
Despite the localized character of this excitation description,
  our result is in relatively good agreement with experiments.

Similar analysis is performed for [Au(tmdt)$_2$].
For [Au(tmdt)$_2$], 
we also find that 
 the inter-molecular transfer  integrals,
 $t_2^{[100]}$, $t_1^{[111]}$, and $t_1^{[001]}$ exceeds the
  intra-molecular transfer integral $t_0$.
These features are qualitatively consistent with those
  obtained by fitting the DFT-based calculation.
 \cite{Seo2008}
In comparison to (TTM-TTP)I$_3$, we found that 
  the two fragments connect by 
the $t_1^{[111]}$ bond form a strong dimer 
($t_1^{[111]}/t_2^{[111]}\simeq 7.0$)
in [Au(tmdt)$_2$], while  
 four fragments interact simultaneously in (TTM-TTP)I$_3$
where  $t_1^{[001]}/t_2^{[001]}\simeq 1.5$.
Due to this feature, the ICO state becomes unfavorable 
    in [Au(tmdt)$_2$].
In addition, it has been pointed out that this system has a good nesting vector
  $q=(1/2,0,0)$ and  
the IAF state with this wave vector 
 is stabilized.\cite{Ishibashi2008,Seo2008}
If we restrict ourselves to the ordering  with wave vector
 $q=(1/2,0,0)$,
  we observe from the strong-coupling analysis,
 that the ICO and CDW states are unstable with respect to the charge-uniform state 
 in which each hole is localized on the $t_1^{[111]}$ bond. 
We note that this charge-uniform state 
is compatible to the IAF state if we take into account the
  antiferromagnetic interactions.
 The energy differences per molecule are
  $E_\mathrm{ICO}-E_\mathrm{IAF}\approx 0.07$ eV and
  $E_\mathrm{CDW}-E_\mathrm{IAF}\approx 0.14$ eV.
More elaborate calculations 
based on our effective model 
should be carried out 
to clarify the origin of a huge magnetic moment 
suggested by the nuclear magnetic resonance measurement.
\cite{Hara2008}

\section{Summary}

In the present paper,
 we have proposed a scheme to determine the parameters of 
  multi-orbital extended Hubbard model from the 
 \textit{ab initio} MR-CI calculations.
To the best of our knowledge,
this is the first theoretical work which aims at
 evaluating  model parameters 
  for a multi-orbital system.
We have applied this method explicitly 
 to the charge-transfer molecular conductor (TTM-TTP)I$_3$ 
and the single-component molecular conductor [Au(tmdt)$_2$].
By taking  advantage of wavefunction-based calculations, 
the CI Hamiltonian matrix 
  for the target model space was constructed, and 
all the model parameters were uniquely determined
 so as to 
  reproduce the different matrix elements.
By examining the band structure,
we have verified the multi-band nature of these systems,
 since the SOMO- and HOMO$-1$-based bands overlap and
these bands  mix in together by the relatively large mixing amplitude.
Furthermore, 
a full fragment decomposition picture 
leading to a parameter hierarchization
has been justified 
 by the observation that 
 the inter-molecular Coulomb repulsions
as well as the intra-molecular interaction $V_0$ 
 follow well the Coulomb $1/r$ law.
Our results strongly support that
the ICO state  experimentally-observed in (TTM-TTP)I$_3$ must be described
 by a  multi-orbital picture.

\acknowledgments

The authors thank S.\ Ishibashi and H.\ Seo 
for the stimulating discussions
at the early stage of the present work.
MT  thanks S.\ Yasuzuka, T.\ Kawamoto, T.\ Mori, and K.\ Yakushi
  for the fruitful discussions on the experimental aspects for the
 TTM-TTP compounds.
MT and YO also thank L.\ Cano-Cort\'es, J. Merino, and K.\ Nakamura
   for discussions
  on the parameters evaluations of molecular solids.
MT was supported by JSPS Institutional Program 
for Young Researcher Overseas Visits.
YO and MLB were supported by the Grant-in-Aid for JSPS Fellows.
This research was also partially supported by
Grant-in-Aid for Scientific Research on Innovative Areas 
(20110002)
from the 
Ministry of Education, Culture, Sports, Science and Technology, Japan.

\end{document}